\documentclass[a4paper]{report}

\usepackage[T1]{fontenc}
\usepackage{textcomp}

\usepackage{amsmath}
\usepackage{amssymb}
\usepackage{amsfonts}
\usepackage{amscd}
\usepackage{bbm}
\usepackage{graphicx}

\usepackage{newcent}

\newcommand{\pd}{\partial}
\newcommand{\de}{\mathrm{d}}
\newcommand{\wec}[1]{\boldsymbol{#1}}
\newcommand{\prp}{{}_{\perp}\!}
\newcommand{\pprp}{{}_{\perp\!\!\!\perp}\!}
\newcommand{\gm}[2]{\left\langle #1,#2 \right\rangle}

\title{Integrability and Chaos -- algebraic and geometric approach}

\author{Tomasz Stachowiak}

\begin{document}



\thispagestyle{empty}

\vspace*{6cm}

\begin{center}

{\LARGE Integrability and Chaos -- algebraic and geometric approach}

\vspace{1.5cm}

{\Large Tomasz Stachowiak}

\vfill

{\large Doctoral thesis written under the supervision of\\
professor Marek Szyd\l owski}

\vspace{1.5cm}
{\large Jagiellonian University, Krak\'ow\\
October 1st, 2008}

\end{center}

\newpage
\thispagestyle{empty}
\vspace*{1cm}
\newpage

\tableofcontents


\newpage
\thispagestyle{empty}
\vspace*{1cm}
\newpage

\chapter{Introduction}


The aim of the present work is to show how the algebraic approach to the
question of integrability can be given geometric foundations. The notion of
first integrals for dynamical systems (Hamiltonian in particular) is almost
always formulated with the tacit assumptions that the underlying phase space
is
Euclidean (or given the additional symplectic structure). The concept of the
metric structure, crucial for Manifolds, is usually omitted -- and no wonder,
since there is no clear way to introduce a distinguished norm for a general
system.

For Lagrangian mechanics there is an equivalent description by means of
the Jacobi or Eisenhart metric with which the flow of the system can be made
geodesic with respect to precisely determined notion of length
\cite{Jacobi,Eisenhart}. However, this
only takes into account the configuration space of the system, not the whole
phase space, which means that for a Hamiltonian system one only considers the
coordinate subspace with second order equations on it and nothing is said about
momenta. This is unacceptable when the characteristic exponents, or chaos is to
be investigated -- the possibly exponential divergence of the trajectories has
to be analysed in the full phase space.

Instead of trying to find one canonical structure, this work presents the
general view that differential geometry has to offer in that topic. A dynamical
system is analysed on a Riemannian manifold, to show how the well-known
equations and definitions have to be modified, with a special view to the
questions of integrability and Lyapunov exponents. The latter are examined in
detail to obtain a differential equation whose solutions are the so called
``time-dependent'' exponents, which tend to the standard ones in infinite time.

The integrability chosen for study here is generally understood to be the
existence of enough first integrals (in involution for the Hamiltonian
systems), which are meromorphic functions over the complexified phase space. On
the one hand, this bears clear consequences on the geometric picture and is
easily translated into the language of differential geometry. On the other, the
question of proving such existence leads to the deeply algebraic properties of
the system such as the analytic continuation of the solutions in complex time,
the differential Galois group and solvability of linear differential equations
by quadratures.

The group-theoretic approach relies on particular solutions and computations of
explicit equations in some coordinate systems. As mentioned before, it is
usually silently assumed from the beginning that the manifold is Euclidean
which leads to significant simplifications when it comes to derivations,
vectors and matrices. Differential geometry requires that we use covariant
derivatives, tensors and distinguish between 1-forms and vectors. It seems like
an unnecessary complication to add but as it turns out, many theorems are much
more straightforward to prove (like the existence of integrals of higher
variational equations for example), some concepts like self-adjoint operators
appear naturally and allow for applications of known theorems from other
branches of mathematics and last but not least, it is possible to identify
which structures are in fact of geometric origin, which can be defined as
coordinate invariant and which are only justified by the efficiency of
calculation.

The main body of this thesis is divided into two parts. In Chapter 2 the
differential Galois group fundamentals are explained, and the basic definitions
and
steps in investigating integrability are described. Next, Chapter 3 deals with
the geometric approach, showing how the concepts introduced earlier need to be
changed, how general dynamical systems are described in this language and
finally showing how Lyapunov exponents can be consistently introduced in a
covariant way and how this leads directly into an easily applicable numerical
routine. Since the notation is index-free (stressing the independence from
coordinate systems or even coordinate bases) almost all derivations are given
in detail. The reason for this is first to make it possible to follow exactly
the flow of exposition if the reader so wishes, and second the fact that there
seem to be very few practical applications of the notation. Hopefully, the
successful formulation of the basics of phase space dynamics presented here 
will prove it is not reserved for pure mathematics only.

The third part of this work contains three examples on applying the Lyapunov
exponents formula or algorithm, and simple comparison of the algebraic
formulation of the normal variational equations versus the same equations as
obtained for the simplest Euclidean manifold with the Levi-Civita connection.

\chapter{Algebraic Approach to Integrability}

Integrability is mostly synonymous with existence of first integrals --
functions that are constant along the solution. However, for some classes of
dynamical systems there are additional, specific requirements.

Consider a general autonomous system
\begin{equation}
    \dot{x_i} = \frac{\de x_i}{\de s} = v_i(x),\quad i=1,\ldots,N.
    \label{gen_dyn_sys}
\end{equation}
It is said to be {\it integrable in the Euler sense} if there exist $N-1$ first
integrals $J_i$, which means that for $i=1,\ldots,N-1$
\begin{equation}
    \dot{J}_i = \frac{\pd J_i}{\pd x_k}v_k = 0.
\end{equation}

Note that only $N-1$ integrals are required, since they formally reduce the
system to a one dimensional one, which can be solved by a quadrature. That last
step gives rise to another constant, which can be considered as the $N$-th
first integral. In the case of autonomous systems this reflects the freedom of
translating the solution ``in time'', or along the trajectory.

Most physical systems posses additional structure of being Hamiltonian. The
dimension is then necessarily even $N=2K$, with the first $K$ coordinates
customarily denoted $q_i$, and the other $K$ -- called momenta -- denoted
$p_i$. The system then has the following form
\begin{equation}
\begin{aligned}
    \dot{q_i} &= \frac{\pd H}{\pd p_i},\\
    \dot{p_i} &= -\frac{\pd H}{\pd q_i},\quad i=1,\ldots,K,
\end{aligned}
\end{equation}
where $H$ is called the Hamiltonian of the system.

While the first integrals are defined exactly as above, these systems are
such a restricted class, that much less is needed for integrability. We say
that s Hamiltonian system is {\it integrable in the Liouville sense} when it
has $K$ first integrals which are in involution. The additional requirement
means that for any two $J_i$ and $J_k$
\begin{equation}
    [J_i,J_k]:=\sum_{j=1}^{K}\left(\frac{\pd J_i}{\pd q_j}
    \frac{\pd J_k}{\pd p_j} -
    \frac{\pd J_i}{\pd p_j}\frac{\pd J_k}{\pd q_j}\right) = 0.
\end{equation}
By construction the Hamiltonian is itself a first integral, and is in
involution with any other additional first integral, as
\begin{equation}
    [J,H] = \dot{J}.
\end{equation}

These are only the basic facts and notation needed here, and a complete
exposition of the topic can be found in \cite{Arnold, Goriely}. The aspect of
algebraic formulation of integrability that is of main interest in this work is
the existence
of first integrals and the criteria or tests of this property. There are no
general algorithms for finding constants of motion explicitly, which goes hand
in hand with the fact that solvable systems are a rare exception (zero measure
set) among all dynamical systems. There are however conditions which integrable
systems must satisfy (i.e. necessary conditions) and that allows of excluding
most systems so that only a few particular cases potentially solvable are left.
They can then be subject to a more detailed analysis not possible for a
whole general class of systems.
The method described in the next section deals with necessary conditions of
existence of first integrals and is one of the most restrictive. It has been
successfully applied to many systems, determining completely the cases which
are non-integrable. For example, it has been proven that a
Hamiltonian system with a homogeneous polynomial potential and natural kinetic
part of any dimension is only integrable in at most finite number of cases
\cite{Marysia}. For other examples of application see 
\cite{Newton,Mgr,Morales_pre}, and
\cite{Morales,Audin} for a detailed introduction.

\section{Differential Galois theory}

The first step required for this approach is to linearise the general equations
\eqref{gen_dyn_sys} around a particular, non-constant solution $\psi(s)$ so
that
\begin{equation}
    x_i(s) = \psi_i(s) + \varepsilon x_i^{(1)} + \varepsilon^2 x_i^{(2)} +\ldots,
\end{equation}
where $\epsilon$ is some small parameter, and each $x^{(n)}$ is called the
$n$-th variation. The original equation then yields in each order of
$\varepsilon$ the $n$-th variational equation
\begin{equation}
    \dot{x}^{(n)}_i = \left.\frac{\pd v_i}{\pd x_j}\right|_{\psi} x^{(n)}_j
    + f_i^{(n)}(\psi,x^{(1)},\ldots,x^{(n-1)}),
\end{equation}
with $f^{(1)}$ identically zero, and in general polynomial in the variations. It is
the first variational equation (VE) that will be of most interest, although the
higher ones are also influenced by integrability providing further necessary
conditions \cite{Morales}.

As the VE is linear it has $N$ linearly independent solutions, and one of them
is simply $\dot{\psi}$. If one imagines that the components of $x^{(1)}$
represent separation of two nearby trajectories of the full system then this
trivial solution corresponds to the displacement relative to the same
trajectory translated in the independent variable $s$. It is called the
tangential part of the VE. Although it is easily solvable on its own, the
complete VE might still as a whole indicate non-integrability. However, there are
$N$ of them, and for the sake of practicality, one is almost often forced to
work with the components left after discarding the tangential part.
This fact gives rise to the normal variational equations.

Formally, the normal bundle is defined to be $\mathrm{N}_{\psi}M :=
\mathrm{T}_{\psi}M/\mathrm{T}\psi$, where for brevity $\psi$ also denotes the
trajectory associated with the particular solution $\psi(s)$, and $M$ is the
base manifold of the original system. The projection
$\pi: \mathrm{T}_{\psi}M\rightarrow\mathrm{N}_{\psi}M$ is then used to
construct the normal variational equations
\begin{equation}
    \dot{y} = \pi\left(\frac{\pd v}{\pd x}\circ\pi^{-1}(y)\right),\quad
    y\in\mathrm{N}_{\psi}M. \label{gen_NVE}
\end{equation}
In this approach, there is no metric structure so, strictly speaking, there is
no orthogonality of
the normal components to the tangent ones. Usually, the particular solution
appears when the first $N-1$ dependent variables $x$ are equal to some constants
and $\psi$ is determined by 1 differential equation. Assuming the Euclidean
structure (of both the base manifold $M$ and thus also of the fibre of
$\mathrm{T}M$), the tangential part of VE is associated with the direction
$x_N$ and the remaining directions are treated as normal.

With Hamiltonian systems, the reduction involves one more step, as there is the
first integral $H$, which can also be used. One more degree of freedom is
eliminated by considering the system \eqref{gen_NVE} on the constant energy
hyper-surface. In terms of the linear variations this means
\begin{equation}
    \de H\circ\pi^{-1}(y) = 0.
\end{equation}
This time the particular solution is sought for so that $N-2$ variables are
(e.g.) zero and a pair of $q_K$ and $p_K$ are left to provide a second order
equation for $\psi$. The above reduction by 2 degrees of freedom then gives the
NVE involving only $q_i$ and $p_i$ for $i=1,\ldots,K-1$ (where, as before,
$N=2K$).

In any case what is left is a set of (non-autonomous) linear differential
equations
\begin{equation}
    \dot{y}=A(s)y, \label{non_aut}
\end{equation}
for which it is possible to define the {\it monodromy group} $\mathcal{M}$.
When
the independent variable is considered as complex, one can ask how the
fundamental matrix of solution changes after analytic continuation of the
solutions in closed loops around a point $s_0$. Since it must be a linear
function of the initial fundamental matrix, a matrix multiplier is obtained,
and after considering all loops, one ends up with a whole group. This subgroup
of $\mathrm{GL}(\mathbb{C})$ will
be an image of the fundamental group $\pi_1(\Psi,s_0)$ of the Riemann surface
$\Psi$ defined by the solution $\psi$.

The existence of a first integrals $J$ of the main system implies there is
also a first integrals of the (normal) variational equations
\begin{equation}
    \left.\frac{\pd J}{\pd x_i}\right|_{\psi}x_i^{(1)} =\mathrm{const},
\end{equation}
provided the gradient does not vanish on the trajectory $\psi$, but even then
the higher derivatives yield a first integral (they cannot all vanish for then
$J$ would trivially be zero). This will be presented in more detail in the next
chapter, as at present it is enough to notice that the above formula gives a
function $f$ such that
\begin{equation}
    f(x^{(1)}) = f(g(x^{(1)})),
\end{equation}
for all $g$ in $\mathcal{M}$. Such $f$ is called a first integral of the
group and it was shown by Ziglin \cite{Ziglin} that if the main system has a
meromorphic first integral then the monodromy group has a rational first
integral. That fact alone can be used for the study of integrability, but since
the above can be refined still, let us pass to the Galois theory itself.

For a linear system \eqref{non_aut}, with the coefficients in some differential
field $K\ni A(s)$ (for example the rational functions of $s$, $\mathbb{C}(s)$,
with the standard derivation of $\de/\de s$), the
solution almost never lies in $K$ but in a larger field $F\supset K$. If the
field extension is generated by all the linearly independent solutions of the
given equation, the field $F$ is called a {\it Piccard-Vessiot extension} of
$K$. It is (up to an isomorphism) unique, and allows to define the {\it
differential Galois group} $\mathcal{G}$ of \eqref{non_aut}, as the group of
automorphisms
of $F$ that leave elements of $K$ fixed and commute with the derivation.

The Galois group is bigger than the monodromy group, and in the special case of
the equations being Fuchsian, $\mathcal{M}$ is dense in $\mathcal{G}$. It is
still an algebraic subgroup of $\mathrm{GL}(K')$, where $K'$ is the field of
constants of $K$ ($\mathbb{C}$ in most applications). It is still the case
that there exist integrals of the Galois group, when there are integrals of the
dynamical system. This leads to two important properties.
\begin{enumerate}
\item
When the system is integrable in the Euler sense with meromorphic first
integrals, the Galois group of the normal variational equations is finite.
\item
For Hamiltonian systems which are integrable in the Liouville sense with
meromorphic first integrals in involution, the identity component of the Galois
group of the (normal) variational equations is abelian.
\end{enumerate}

The above are fundamental theorems of the theory, but for practical
applications one is usually interested in the consequences for the solutions of
the NVE. In the former case the solutions lie in an algebraic extension of the
field $K$, and in the latter in a (generalised) {\it Liouvillian extension}.
Such an extension is a formally defined concept of a solution in a ``closed
form'' and arises from $K$ after finitely many steps each of which consists of
including one of three types of a new element: one that is algebraic over $K$,
its derivative lies in $K$ or the derivative of its logarithm lies in $K$.

In physical applications, the NVE are usually of the order two, and there is a
general tool for checking the above property: the Kovacic algorithm
\cite{Kovacic}. There are also particular results for special families of
equations like the Riemann P-equation \cite{Kimura} or Lam\'e equation
\cite{Beukers}, which give explicit conditions on the equations' parameters for
their solution to be Liouvillian. The examples cited in the previous section
show how effective the method is -- reducing a general problem of the existence
of first integrals to the question of solvability of linear differential
equations.

Before the other (geometric) description is presented, one remark is in order.
Although the Galois group approach is applied to physical systems, for which
one naturally considers the variables to be real, the mathematical ``workshop''
is located in the complex domain. This means that not only the dependent
variables but also the time itself can be imaginary, and this has several
implications. First of all, for all physical systems where one assumes
sufficient smoothness, the lack of complex integrals implies the lack of real
ones (as a special case). However, there are systems for which there exist real
smooth integrals, that are not even real-analytic \cite{Gorni}. Thus,
integrability is always depends on the domain considered, and the family of
functions considered ``good enough'' to be first integrals.

Secondly, some systems although non-integrable behave chaotically in the
imaginary phase space and not in the real one. A particular example of that is
the Gross-Neveu system analysed in \cite{GN}, which will also be studied here
with the help of Lyapunov exponents. Since the algebraic theory is usually
studied without much input of differential geometry (as there is usually no
need for metric structure) the exact connection between integrability and
geometric concepts like exponential separation of trajectories or topological
chaos remains elusive. The next chapter is thus devoted to treating dynamical
system with more detailed differential geometry and, as an example, giving a
new formulation of the Lyapunov exponents together with some implications that
first integrals enforce and practical results.

\chapter{Geometric View}

Throughout this chapter the index-free, operator notation will be used, and the
convention is mostly that of \cite{Skwarczynski} or \cite{Kobayashi} -- the
reader can find in depth introduction to the subject in those two books.

Vectors
and 1-forms will be denoted by bold symbols, and $\wec{A}$, $\wec{B}$,
$\wec{X}$, $\wec{X}_1$, $\wec{X}_2$ and so on will mean arbitrary vectors and
$\wec{\phi}$ will signify any 1-form (usually used to define an operator or
prove an equality). Vectors will be considered to act on functions as
differential operators so that in a coordinate map $u$
\begin{equation}
    \wec{X}(f) = \sum_{i=1}^{N}X^i\frac{\pd f}{\pd u^i},
\end{equation}
where $N$ is the dimension of the base manifold.

The covariant derivative $\nabla$ is defined for $\binom{1}{k}$ type tensors
(the exterior tangent bundle strictly speaking) by the property
\begin{equation}
    \nabla(\beta\otimes\wec{X}) = \de\beta\otimes\wec{X}+
    (-1)^k\beta\wedge\nabla\wec{X},
\end{equation}
with $\beta$ being a $k$-form (or a function for $k=0$). And the connection
itself is fully characterised by its action on any basis $\{\wec{E}_i\}$
\begin{equation}
    \nabla\wec{E_i} = \sum_j\Omega_{ij}\otimes\wec{E}_j,
\end{equation}
where $\Omega$ is a matrix of 1-forms (not a tensor). It allows to define the
torsion tensor, the curvature endomorphism and the Ricci tensor as
\begin{equation}
\begin{aligned}
    T(\wec{X}_1,\wec{X}_2) &:= \nabla_{\wec{X}_1}\wec{X}_2 -
    \nabla_{\wec{X}_2}\wec{X}_1 - [\wec{X}_1,\wec{X}_2],\\
    R(X_1,X_2)\wec{A} &:= (\nabla_{\wec{X}_1}\nabla_{\wec{X}_2}-
    \nabla_{\wec{X}_2}\nabla_{\wec{X}_1} -
    \nabla_{[\wec{X}_1,\wec{X}_2]})\wec{A}\\
    \mathcal{R}(\wec{A},\wec{B}) &:=
    \mathrm{tr}(\wec{X}\rightarrow R(\wec{A},\wec{X})\wec{B}).
\end{aligned}
\end{equation}

With the introduction of the metric structure (a symmetric tensor)
$\mathfrak{g}$
\begin{equation}
    \mathfrak{g}(\wec{X}_1,\wec{X}_2) \rightarrow 
    \gm{\wec{X}_1}{\wec{X}_2}\in\mathbb{R},
\end{equation}
the Levi-Civita (torsionless and Riemannian) connection can be introduced by
taking
\begin{equation}
\begin{aligned}
    T &= 0,\\
    \de\gm{\wec{X}_1}{\wec{X}_2} &= 
    \gm{\nabla\wec{X}_1}{\wec{X}_2} + \gm{\wec{X}_1}{\nabla\wec{X}_2}.
\end{aligned}
\end{equation}
The second requirement implies also
\begin{equation}
    \de\eta(\wec{X}_1,\ldots,\wec{X}_N) = 
    \sum_j\eta(\wec{X}_1,\ldots,\nabla\wec{X}_j,\ldots,\wec{X}_N),
    \label{par_vol}
\end{equation}
for the volume form $\eta$. The metric also allows to define the Riemann
tensor
\begin{equation}
    R(\wec{X}_1,\wec{X}_2,\wec{A},\wec{B}) =
    \gm{R(\wec{X}_1,\wec{X}_2)\wec{A}}{\wec{B}}.
\end{equation}

The interior product $\iota$ is defined as
\begin{equation}
    (\iota_{\wec{X}}\zeta)(\wec{X}_1,\ldots,\wec{X}_{k}) := 
    \zeta(\wec{X},\wec{X}_1,\ldots,\wec{X}_k),
\end{equation}
where $\zeta$ is some $(k+1)$-form. The following property of the Lie
derivative $\pounds$ will also be useful at some point
\begin{equation}
    \pounds_{\wec{X}}\zeta = \de(\iota_{\wec{X}}\zeta) 
    + \iota_{\wec{X}}(\de\zeta).
\end{equation}

\section{The variational equations}

\begin{figure}[h]
\includegraphics[width=\textwidth]{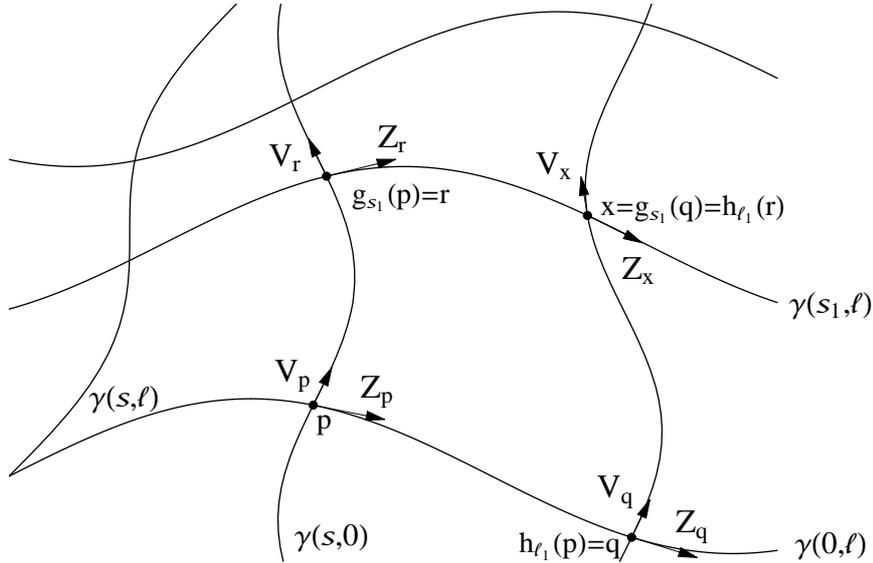}
\caption{A visualisation of the family of curves $\gamma(s,l)$ generated by the
fields $\wec{V}$ and $\wec{Z}$ (with their dependence on the point indicated).}
\label{flows_vz}
\end{figure}

Consider now the following construction (Figure \ref{flows_vz}). Given the
field $\wec{V}$ and a curve $\gamma(0,l)$ (not being one of the field's
integral curves), we can construct the image of $\gamma(l)$ under the flow of
$\wec{V}$ with $s=s_1$. The image curve will be denoted by $\gamma(s_1,l)$, and
by using the same $l$ we understand, that each point of the new
curve is the image of a point of the original curve for that value of $l$.
Thus, the parameter $l$, which needs not be the natural parameter, gives rise to
a new vector field $\wec{Z} := \frac{\pd}{\pd l}$ tangent to family of
transformed curves (along with the original one).

By construction, for any two points $p$ and $q$ on $\gamma(0,l)$ carried into
$r$ and $x$ on
$\gamma(s_1,l)$ the differences of their $l$ values are the same. Or, in other
words, that the flows generated by $\wec{V}$ and $\wec{Z}$ ($g$ and $h$
respectively) commute
\begin{equation}
    g_{s_1}(h_{l_1}(p)) = h_{l_1}(g_{s_1}(p)) \iff [\wec{V},\wec{Z}]=0.
\end{equation}
Taking into account that the connection is torsion-free, this will allow us to
transform the derivative of $\wec{Z}$ along $\gamma(s)$.
\begin{equation}
    T(\wec{V},\wec{Z}) -
    [\wec{V},\wec{Z}] = 
    \nabla_{\wec{V}}\wec{Z} - \nabla_{\wec{Z}}\wec{V} = 0
    \quad \Rightarrow \quad \frac{D}{\pd s}\wec{Z} := 
    \nabla_{\wec{V}}\wec{Z} = \nabla_{\wec{Z}}\wec{V}, \label{comm}
\end{equation}
which is, in fact, the first variational equation (VE)
\begin{equation}
    \dot{\wec{Z}} = (\nabla\wec{V})(\wec{Z}),
\end{equation}
where the dot will, from now on, denote the covariant derivative along the
field $\wec{V}$.

\section{Projected variational and deviation equations}
Let us introduce the operator $h$ projecting a vector field on the normal
(with respect to $\wec{V}$) bundle $\mathrm{N}M$
\begin{equation}
    h = \boldsymbol 1 - \wec{\upsilon}\otimes\wec{V},
\end{equation}
where $\wec{\upsilon}$ is a one-form dual to $\wec{V}$ given by
\begin{equation}
    \wec{\upsilon}(\wec{X}) = 
    \frac{\langle\wec{V},\wec{X}\rangle}{\langle\wec{V},\wec{V}\rangle},
\end{equation}
so that $\wec{\upsilon}(\wec{V})=1$.
Obviously we have
\begin{equation}
    h^2 = h,\qquad h(\wec{V})=0,
\end{equation}
and for constant norm of $\wec{V}$
\begin{equation}
    \gm{\nabla\wec{V}}{\wec{V}} = 0\;\Rightarrow\;
    \wec{\upsilon}(\nabla\wec{V}) = 0.
\end{equation}
The norm will also be written shortly as
$1/\mathcal{N}:= \langle\wec{V},\wec{V}\rangle$, and will be included
explicitly in calculations so that all the formulae hold also in the
Lorentzian case $\mathcal{N}=-1$. 
For brevity, a projected vector will also be
denoted by \[ \prp\wec{Z}:=h(\wec{Z}). \]

Consider now the derivative of the projection along the curve $\gamma(s)$
\begin{equation}
\begin{aligned}
    \frac{D}{\pd s}\prp\wec{Z}
    &= \nabla_{\wec{V}}(\wec{Z}-\wec{\upsilon}(\wec{Z})\wec{V})\\
    &= \nabla_{\wec{Z}}\wec{V} -\wec{\upsilon}(\wec{Z})\nabla_{\wec{V}}\wec{V}-
    (\nabla_{\wec{V}}\wec{\upsilon}(\wec{Z}))\wec{V}\\
    &= \nabla_{(\wec{Z}-\wec{\upsilon}(\wec{Z})\wec{V})}\wec{V} -
    (\nabla_{\wec{V}}\wec{\upsilon}(\wec{Z}))\wec{V}\\
    &= \nabla_{\prp\wec{Z}}\wec{V} -
    \wec{V}(\wec{\upsilon}(\wec{Z}))\wec{V}, \label{full_proj}
\end{aligned}
\end{equation}
where we have used the commutation property (\ref{comm}). Acting with $h$
on both sides of the above equality we obtain
\begin{equation}
     \prp\frac{D}{\pd s}\prp\wec{Z} = \prp\nabla_{\prp\wec{Z}}\wec{V}. \label{der1}
\end{equation}
To obtain the deviation equation we differentiate again
\begin{equation}
\begin{aligned}
    \frac{D}{\pd s}\prp\frac{D}{\pd s}\prp\wec{Z}
    &= \nabla_{\wec{V}}\prp\nabla_{\prp\wec{Z}}\wec{V}\\
    &= \nabla_{\wec{V}}\nabla_{\prp\wec{Z}}\wec{V}-
       \nabla_{\wec{V}}(\wec{\upsilon}(\nabla_{\prp\wec{Z}}\wec{V})\wec{V})\\
    &= R(\wec{V},\prp\wec{Z})\wec{V}+\nabla_{\prp\wec{Z}}\dot{\wec{V}}+
    \nabla_{[\wec{V},\prp\wec{Z}]}\wec{V},
\end{aligned}
\end{equation}
and project onto $\mathrm{N}M$
\begin{equation}
\begin{aligned}
    \prp\frac{D}{\pd s}\prp\frac{D}{\pd s}\prp\wec{Z}
    &= h\left(R(\wec{V},\prp\wec{Z})\wec{V}\right)
    +\prp\nabla_{\prp\wec{Z}}\dot{\wec{V}}+
    \prp\nabla_{[\wec{V},\prp\wec{Z}]}\wec{V}\\
    &= R(\wec{V},\prp\wec{Z})\wec{V} - \mathcal{N}
    \langle\wec{V},R(\wec{V},\prp\wec{Z})\wec{V}\rangle + 
    \prp\nabla_{\prp\wec{Z}}\dot{\wec{V}} +
    \prp\nabla_{[\wec{V},\prp\wec{Z}]}\wec{V}\\
    &= R(\wec{V},\prp\wec{Z})\wec{V} + 
    \prp\nabla_{\prp\wec{Z}}\dot{\wec{V}} +
    \prp\nabla_{[\wec{V},\prp\wec{Z}]}\wec{V}.
\end{aligned}
\label{proj2der}
\end{equation}
The commutator can be further simplified
\begin{equation}
\begin{aligned}
    \left[\wec{V},\prp\wec{Z}\right]
    &= [\wec{V},\wec{Z}-\wec{\upsilon}(\wec{Z})\wec{V}]\\
    &= [\wec{V},-\wec{\upsilon}(\wec{Z})\wec{V}]\\
    &= -\wec{V}\left(\wec{\upsilon}(\wec{Z})\right)\wec{V}
    -\wec{\upsilon}(\wec{Z})[\wec{V},\wec{V}]\\
    &= -\mathcal{N}\wec{V}\left(\gm{\wec{V}}{\wec{Z}}\right)\wec{V}\\
    &= -\mathcal{N}\left(\gm{\dot{\wec{V}}}{\wec{Z}}
    +\gm{\wec{V}}{\nabla_{\wec{Z}}\wec{V}}\right)\wec{V}\\
    &= -\mathcal{N}\gm{\dot{\wec{V}}}{\prp\wec{Z}}\wec{V},
\end{aligned}
\end{equation}
the scalar coefficient above is
\begin{equation}
\begin{aligned}
    \wec{V}(\wec{\upsilon}(\wec{Z}))
    &= \mathcal{N}(\gm{\dot{\wec{V}}}{\wec{Z}}+\gm{\wec{V}}{\dot{\wec{Z}}})
    -2\mathcal{N}^2\gm{\wec{V}}{\dot{\wec{V}}} \gm{\wec{V}}{\wec{Z}}\\
    &= \mathcal{N}\left(\gm{\dot{\wec{V}}}{\prp\wec{Z}}+
    \gm{\dot{\wec{V}}}{\wec{V}}\wec{\upsilon}(\wec{Z})+
    \gm{\wec{V}}{\nabla_{\wec{Z}}\wec{V}}
    -2\gm{\wec{V}}{\dot{\wec{V}}}\wec{\upsilon}(\wec{Z})\right)\\
    &= \mathcal{N}\left(\gm{\dot{\wec{V}}}{\prp\wec{Z}}
    -\gm{\dot{\wec{V}}}{\wec{V}}\wec{\upsilon}(\wec{Z})
    +\gm{\wec{V}}{\nabla_{\prp\wec{Z}}\wec{V}}
    +\wec{\upsilon}(\wec{Z})\gm{\wec{V}}{\dot{\wec{V}}}\right)\\
    &= \mathcal{N}(\gm{\dot{\wec{V}}}{\prp\wec{Z}}
    +\gm{\wec{V}}{\nabla_{\prp\wec{Z}}\wec{V}}).
\end{aligned}
\end{equation}
Substituting the above, the second derivative becomes
\begin{equation}
    \prp\frac{D}{\pd s}
    \prp\frac{D}{\pd s}\prp\wec{Z}=
    R(\wec{V},\prp\wec{Z})\wec{V} +
    \prp\nabla_{\prp\wec{Z}}\dot{\wec{V}}
    -\mathcal{N}\gm{\prp\dot{\wec{V}}}{\prp\wec{Z}}\prp\dot{\wec{V}}.
    \label{der2}
\end{equation}

This can be written in a shorter form, which
allows further analysis to be simpler, upon introducing the Fermi derivative.
Since $\wec{V}$
does not necessarily define a geodesic flow, we want to define a new derivation
along $\gamma(s)$ which would satisfy the following properties:
\begin{enumerate}
\item $\dfrac{D_F}{\pd s}\wec{V} = 0$.
\item $\dfrac{D_F}{\pd s}\wec{X}=\dfrac{D}{\pd s}\wec{X}$, for geodesic
$\wec{V}$.
\item If $\dfrac{D_F}{\pd s}\wec{X}=0=\dfrac{D_F}{\pd s}\wec{U}$, then
$\dfrac{\de}{\de s}\langle\wec{X},\wec{U}\rangle = 0$.
\item $\dfrac{D_F}{\pd s}\wec{X} =\prp\dfrac{D}{\pd s}\wec{X}$,
for $\wec{X}$ orthogonal to $\wec{V}$.
\end{enumerate}
The last property allows us to find the explicit form of the Fermi derivative
\begin{equation}
    \frac{D_F}{\pd s}\wec{X} = \frac{D}{\pd s}\wec{X} 
    -\mathcal{N}\gm{\wec{V}}{\wec{X}}\dot{\wec{V}}
    +\mathcal{N}\gm{\dot{\wec{V}}}{\wec{X}}\wec{V}.
\end{equation}
Equations (\ref{der1}) and (\ref{der2}) now read
\begin{equation}
\begin{aligned}
    \frac{D_F}{\pd s}\prp\wec{Z} &= \prp\nabla_{\prp\wec{Z}}\wec{V}
    =:\prp\mathcal{A}(\prp\wec{Z}),\\
    \frac{D_F^2}{\pd s^2}\prp\wec{Z} &= 
    R(\wec{V},\prp\wec{Z})\wec{V} +
    \prp\nabla_{\prp\wec{Z}}\dot{\wec{V}}
    -\mathcal{N}\gm{\prp\wec{Z}}{\prp\dot{\wec{V}}}\prp\dot{\wec{V}}
    =:\Phi(\prp\wec{Z}).
\end{aligned}
\end{equation}
These can be called the projected variational and deviation equation
respectively. For both operators introduced here we have
\begin{equation}
    \prp\mathcal{A},\;\Phi: \mathrm{T}M \rightarrow \mathrm{N}M,
\end{equation}
where the normal bundle is taken with respect to $\wec{V}$,
so that the matrix of components of $\prp\mathcal{A}$ explicitly only has
non-zero elements for directions orthogonal to $\wec{V}$, and is effectively
a $(N\!-\!1)\times(N\!-\!1)$ matrix (when acting on $\mathrm{N}M$).

\section{The Raychaudhuri equation}

Usually one decomposes $\mathcal{A}$ into its
symmetric and antisymmetric parts using the corresponding index notation
\begin{equation}
    V_{\mu;\nu} = V_{(\mu;\nu)} + V_{[\mu;\nu]}.
\end{equation}
However, with the assumed definition of the covariant derivative, $\mathcal{A}$
is not a tensor of type $\binom{0}{2}$, but can rather be identified with a
tensor of type $\binom{1}{1}$. The question of ``transposing'' such an operator
can be dealt with naturally, when one recalls that the metric tensor defines
the musical isomorphism, so that the following diagram commutes
\begin{equation}
\begin{CD}
    \mathcal{A} @>\mathfrak{g}>> \flat\mathcal{A} \\
    @V\dag VV  @VV T V \\
    \mathcal{A}^{\dag}
    @<\mathfrak{g}^{-1}<< (\flat\mathcal{A})^T
\end{CD},
\end{equation}
where the new symbols above are defined as
\begin{equation}
\begin{aligned}
    (\flat{A})(\wec{X},\wec{U}) =
    \langle A(\wec{X}),\wec{U}\rangle,\\
    (\wec{\phi}^T)(\wec{X},\wec{U}) = \wec{\phi}(\wec{U},\wec{X}),\\
    \langle(\sharp\wec{\phi})(\wec{X}),\wec{U}\rangle =
    \wec{\phi}(\wec{X},\wec{U}),
\end{aligned}
\end{equation}
It is, of course, possible to only define the \textit{Hermitian adjoint
operator} of $A$ as
\begin{equation}
    \langle A^{\dag}(\wec{X}),\wec{U}\rangle :=
    \langle \wec{X}, A(\wec{U})\rangle,
\end{equation}
but then the analogy with simple transposition is not as clearly visible. As
can be seen the adjoint is really a transposition of arguments in the 2-form
$\phi$ canonically isomorphic to the given endomorphism $A$. In the index
notation the above operations take the form
\begin{equation}
\begin{aligned}
    (\flat A)_{\mu\nu} =
    \mathfrak{g}_{\mu\sigma}A^{\sigma}_{\phantom{\sigma}\nu},&&
    (\wec{\phi}^T)_{\mu\nu} = \phi_{\nu\mu},\\
    (\sharp\wec{\phi})^{\mu}_{\phantom{\mu}\nu} =
    \mathfrak{g}^{\mu\sigma}\phi_{\sigma\nu}, &&
    (A^{\dag})^{\mu}_{\phantom{\mu}\nu} = A_{\nu}^{\phantom{\nu}\mu}.
\end{aligned}
\end{equation}

The adjoint is especially simple for operators of the form
$\wec{\phi}\otimes\wec{X}$ as can be directly checked
\begin{equation}
\begin{aligned}
    \gm{\wec{A}}{(\wec{\phi}\otimes\wec{X})(\wec{B})}
    &=\gm{\wec{A}}{\wec{\phi}(\wec{B})\wec{X}}\\
    &=\wec{\phi}(\wec{B})\gm{\wec{A}}{\wec{X}}\\
    &=\flat\wec{X}(\wec{A})\gm{\sharp\wec{\phi}}{\wec{B}}\\
    &=\gm{(\flat\wec{X}\otimes\sharp\wec{\phi})(\wec{A})}{\wec{B}},
\end{aligned}
\label{sim_ad}
\end{equation}
or $(\wec{\phi}\otimes\wec{X})^{\dag}
=(\flat\wec{X})\otimes(\sharp\wec{\phi})$.

We are now ready to decompose $\prp\nabla\wec{V}$ into its self-adjoint
(Hermitian) and anti-self-adjoint (anti-Hermitian) parts. This is essentially
the symmetric splitting of the associated 2-form $\phi$. The reason
for carrying out this procedure in the operator approach is that we are dealing
with linear differential equations and any eigenvalue problem will be much more
straightforward.
We will be able to use all the standard theorems regarding Hermitian operators.

Decomposing now with regard to $\dag$ we have
\begin{equation}
    \frac{D_F}{\pd s}\wec{\prp Z} =
    \theta(\prp\wec{Z}) + \omega(\prp\wec{Z}),
\end{equation}
with
\begin{equation}
\begin{aligned}
    \theta &= \prp\mathcal{A}^H 
    := \frac12(\prp\mathcal{A}+\prp\mathcal{A}^{\dag}),\\
    \omega &= \prp\mathcal {A}^A 
    := \frac12(\prp\mathcal{A}-\prp\mathcal{A}^{\dag}),
\end{aligned}
\end{equation}
so that
\begin{equation}
    \theta^{\dag} = \theta,\quad \omega^{\dag} = -\omega.
\end{equation}
One could think of $\theta$ and $\omega$ as generating the flow's expansion and
rotation respectively.

We turn next to the deviation equation to obtain the derivative of
$\prp\mathcal{A}$
\begin{equation}
    \frac{D^2_F}{\pd s^2}\prp\wec{Z}=\Phi(\prp\wec{Z})=
    \frac{D_F}{\pd s}(\prp\mathcal{A}(\wec{Z})) =:
    \frac{D_F\prp\mathcal{A}}{\pd s}(\prp\wec{Z})+
    \prp\mathcal{A}\left(\frac{D_F}{\pd s}\prp\wec{Z}\right),\label{op_der}
\end{equation}
which we rewrite as
\begin{equation}
\begin{aligned}
    \frac{D_F\prp\mathcal{A}}{\pd s} &= \frac{D_F\theta}{\pd s} +
    \frac{D_F\omega}{\pd s} = \Phi - \prp\mathcal{A}^2,\\
    &= \Phi - \theta^2 - \omega^2 -\{\theta,\omega\},
\end{aligned}
\label{deriv_A}
\end{equation}
where $\{,\}$ stand for the anti-commutator. It is straightforward to construct
the Hermitian decomposition of the left-hand side of the above equation, but
one must ask if the derivatives of $\theta$ and $\omega$ are themselves the
\mbox{(anti-)}Hermitian parts of the left-hand side. Or, in other words, if
\begin{equation}
    \frac{D_F A^{\dag}}{\pd s} \overset{?}{=}
    \left(\frac{D_F A}{\pd s}\right)^{\dag}.
\label{is_F}
\end{equation}
First we prove this holds for the covariant derivative (which is expected for a
Riemannian connection)
\begin{equation}
\begin{aligned}
    \gm{\left(\frac{DA}{\pd s}\right)^{\dag}\wec{X}}{\wec{U}}
    &=\gm{\wec{X}}{\frac{DA}{\pd s}(\wec{U})}\\
    &=\gm{\wec{X}}{\frac{D}{\pd s}A(\wec{U})} 
    - \gm{\wec{X}}{A\left(\frac{D\wec{U}}{\pd s}\right)}\\
    &=\wec{V}(\gm{\wec{X}}{A(\wec{U})})
    -\gm{\frac{D\wec{X}}{\pd s}}{A(\wec{U})}
    -\gm{A^{\dag}(\wec{X})}{\frac{D\wec{U}}{\pd s}}\\
    &=\wec{V}(\gm{A^{\dag}(\wec{X})}{\wec{U}})
    -\gm{A^{\dag}\left(\frac{D\wec{X}}{\pd s}\right)}{\wec{U}}\\
    &\phantom{=}-\wec{V}(\gm{A^{\dag}(\wec{X})}{\wec{U}})
    +\gm{\frac{D}{\pd s}A^{\dag}(\wec{X})}{\wec{U}}\\
    &=\gm{\frac{D}{\pd s}A^{\dag}(\wec{X})
    - A^{\dag}\left(\frac{D\wec{X}}{\pd s}\right)}{\wec{U}}\\
    &=\gm{\frac{DA^{\dag}}{\pd s}\wec{X}}{\wec{U}}.
\end{aligned}
\end{equation}
For the Fermi derivative it thus suffices to check the behaviour of the last
two terms, which are all of the form \eqref{sim_ad}. They
can be written as
\begin{equation*}
    W_1=-\mathcal{N}(\flat\wec{V}\otimes\dot{\wec{V}}-
    \flat\dot{\wec{V}}\otimes\wec{V}),
\end{equation*}
so one immediately has $W_1^{\dag}=-W_1$. The question \eqref{is_F} can now be
rephrased as
\begin{equation}
    (W_1\circ A - A\circ W_1)^{\dag} \stackrel{?}{=}
    W_1\circ A^{\dag} - A^{\dag}\circ W_1.
\end{equation}
As $ (A\circ B)^{\dag} = B^{\dag}\circ A^{\dag}$, the above is identically
satisfied for this particular $W_1$.

As we assume zero torsion, and the connection to be
Riemannian, we have for the Riemann tensor
\[R(A,B,X,Y) = R(X,Y,A,B)\] so that using an auxiliary operator
\[W_R(\wec{X})=R(\wec{V},\wec{X})\wec{V}\] it is possible to obtain the
adjoint of $R$ as a function of its second argument
\begin{equation}
\begin{aligned}
    \langle W_R(\wec{X}),\wec{U}\rangle &=
    R(\wec{V},\wec{U},\wec{V}\wec{X}) :=
    \langle R(\wec{V},\wec{U})\wec{V},\wec{X}\rangle \\
    &= \langle \wec{X},R(\wec{V},\wec{U})\wec{V}\rangle\\
    &= \langle \wec{X},W_R^{\dag}(\wec{U})\rangle =
    \langle \wec{X},W_R(\wec{U})\rangle,
\end{aligned}
\end{equation}
and it turns out to be self-adjoint. This allows us to write
\begin{equation}
    \Phi^{\dag} =
    W_R+
    (\prp\nabla\dot{\wec{V}})^{\dag} -
    \mathcal{N}\flat\prp\dot{\wec{V}}\otimes\prp\dot{\wec{V}},
\end{equation}
or
\begin{equation}
\begin{aligned}
    \Phi^H &= W_R + (\prp\nabla\dot{\wec{V}})^H -
    \mathcal{N}(\flat\prp\dot{\wec{V}})\otimes\prp\dot{\wec{V}},\\
    \Phi^A &= (\prp\nabla\dot{\wec{V}})^A.
\end{aligned}
\end{equation}
Together with equation (\ref{deriv_A}) this gives
\begin{equation}
\begin{aligned}
    \frac{D_F\theta}{\pd s} &=
    W_R + (\prp\nabla\dot{\wec{V}})^H -
    \mathcal{N}(\flat\prp\dot{\wec{V}})\otimes\prp\dot{\wec{V}} 
    -\theta^2-\omega^2,\\
    \frac{D_F\omega}{\pd s} &=
    (\prp\nabla\dot{\wec{V}})^A - \{\theta,\omega\}.
\end{aligned}
\label{th_om_derivs}
\end{equation}
We introduce here a new quantity $\vartheta = \mathrm{tr}\,\theta$
\begin{equation}
    \mathrm{tr}\frac{D_F\theta}{\pd s} =
    \frac{D_F}{\pd s}(\mathrm{tr}\,\theta) = \dot{\vartheta},
\end{equation}
since the Fermi derivative of a scalar function is just the derivative with
respect to $s$. Taking the trace of the right-hand side of the first of
equations (\ref{th_om_derivs}) yields
\begin{equation}
\begin{aligned}
    \mathrm{tr}\frac{D_F\theta}{\pd s} & =
    \mathcal{R}(\wec{V},\wec{V}) + \mathrm{tr}(\prp\nabla\dot{\wec{V}})
    -\mathcal{N}\gm{\dot{\wec{V}}}{\dot{\wec{V}}}
    -\mathrm{tr}(\theta^2+\omega^2)\\ &=
    \mathcal{R}(\wec{V},\wec{V}) + \mathrm{tr}(\nabla\dot{\wec{V}} 
    -\mathcal{N}\gm{\nabla\dot{\wec{V}}}{\wec{V}}\otimes\wec{V})
    -\mathcal{N}\gm{\dot{\wec{V}}}{\dot{\wec{V}}}
    -\mathrm{tr}(\theta^2+\omega^2)\\ &=
    \mathcal{R}(\wec{V},\wec{V}) + \mathrm{tr}(\nabla\dot{\wec{V}})
    +\mathcal{N}\gm{\dot{\wec{V}}}{\nabla_{\wec{V}}\wec{V}}
    -\mathcal{N}\gm{\dot{\wec{V}}}{\dot{\wec{V}}}
    -\mathrm{tr}(\theta^2+\omega^2)\\ &=
    \mathcal{R}(\wec{V},\wec{V}) + \mathrm{tr}(\nabla\dot{\wec{V}})
    -\mathrm{tr}(\theta^2+\omega^2),
\end{aligned}
\end{equation}
where we have used the fact that \[\nabla\gm{\dot{\wec{V}}}{\wec{V}}=0\] and
that \[\mathrm{tr}(\wec{\phi}\otimes\wec{X}) = \wec{\phi}(\wec{X}).\]
Writing conventionally 
\[\theta = \sigma + \frac{\vartheta}{N\!-\!1}\mathbbm{1}\]
to separate the trace-free part $\sigma$, one finally arrives at the
Raychaudhuri equation
\begin{equation}
    \frac{\de}{\de s}\vartheta = \mathcal{R}(\wec{V},\wec{V}) -
    \mathrm{tr}\,\omega^2 - \mathrm{tr}\,\sigma^2
    - \frac{1}{N\!-\!1}\vartheta^2 + \mathrm{tr}(\nabla\dot{\wec{V}}).
    \label{Ray}
\end{equation}

\section{Higher order variational equations}

Let us turn now to the flows generated by the fields $\wec{V}$ and $\wec{Z}$ as
shown on Figure \ref{flows_vz}. Let $u:M\rightarrow\mathbb{R}^N$ denote a local
set of coordinates, and
consider a function $f$ at point $r=g_s(p)$, where the local diffeomorphism
$g_s$ is associated to the vector field $\wec{V}$ so that
\begin{equation}
    \frac{\de}{\de s} u(g_s(p)) =: \wec{V}_r(u) = f_r.
\end{equation}
These are simply the field's components in the coordinate basis associated with
$u$ (not that the subscript indicates the point, not a vector component).
Taking their derivative with respect to $s$ we write
\begin{equation}
    \frac{\de}{\de s} f_r = \wec{V}_r(f),
\end{equation}
or
\begin{equation}
    \frac{\de^2}{\de s^2} u(g_s(p)) = \wec{V}_r\left(
    \wec{V}_r(u)\right),
\end{equation}
so that
\begin{equation}
    \frac{\de^n}{\de s^n} u(g_s(p)) = \wec{V}_r^{(n)}(u).
\end{equation}
Take now another point $q=h_l(p)$, generated with the flow of $\wec{Z}$, and
expand its coordinates, as is always possible in a local map
\begin{equation}
\begin{aligned}
    u(q) = u(h_l(p)) &= \sum_{n=0}^{\infty}\frac{l^n}{n!}\left[
    \frac{\de^n}{\de l^n} u(h_l(p))\right]_{l=0}\\
    &= \sum_{n=0}^{\infty}\frac{l^n}{n!}\left[
    \wec{Z}^{(n)}_{h_l(p)}u(h_l(p))\right]_p\\
    &= \sum_{n=0}^{\infty}\frac{l^n}{n!}
    \wec{Z}^{(n)}_p(u)=\exp(l\wec{Z})_p(u). \label{Taylor}
\end{aligned}
\end{equation}
We define the sequence of displacements $u_n$ to be $\wec{Z}_p^{(n)}(u)$, and
call it variations of the $n$-th order. The partial sums correspond to points
$p_n$ such that $u(p_n) = \sum_{i\leq n} u_i$, and the $n$-th variation is then
a vector in $\mathbb{R}^N$ (coordinate space) connecting the point $p_{n-1}$ to
$p_n$.

Thus, only one vector field is needed to describe all the variations, although
we can formally write the higher order equation (HVE) as
\begin{equation}
    \frac{\de}{\de s}u_n = \wec{V}\left(\wec{Z}^{(n)}(u)\right) =
    \wec{Z}^{(n)}\left(\wec{V}(u)\right),
\end{equation}
by analogy to the original dynamical system
\[\dot{u} = \wec{V}(u) \iff \dot{u}^{\mu} = V^{\mu}.\]
For example, the first VE equation is
\begin{equation}
\begin{aligned}
    \frac{\de}{\de s}\left(\wec{Z}(u^{\mu})\right) 
    &= \wec{Z}(\wec{V}(u^{\mu})) = \wec{Z}(V^{\mu}),\\
    \dot{Z}^{\mu} &= Z^{\nu} \frac{\pd V^{\mu}}{\pd u^{\nu}},
\end{aligned}
\end{equation}
and the second
\begin{equation}
\begin{aligned}
    \frac{\de}{\de s}\left(\wec{Z}(\wec{Z}(u^{\mu}))\right) 
    &= \wec{Z}(\wec{Z}(V^{\mu})),\\
    \frac{\de}{\de s}
    \left(Z^{\nu}\frac{\pd Z^{\mu}}{\pd u^{\nu}}\right) 
    &= Z^{\nu}Z^{\lambda} 
    \frac{\pd^2 V^{\mu}}{\pd u^{\nu}\pd u^{\lambda}} + 
    Z^{\nu}\frac{\pd Z^{\lambda}}{\pd u^{\nu}} \frac{\pd V^{\mu}}{\pd
    u^{\lambda}},
\end{aligned}
\end{equation}
or, in the usual notation,
\begin{equation}
\begin{aligned}
    \dot{u}_1^{\mu} &= u_1^{\nu} \pd_{\nu} V^{\mu} \\
    \dot{u}_2^{\mu} &= u_1^{\nu}u_1^{\lambda}\pd_{\nu\lambda}^2 V^{\mu}+ 
    u_2^{\nu} \pd_{\nu}V^{\mu}.
\end{aligned}
\end{equation}

The whole construction is evidently coordinate-dependent, as the chart $u$
enters explicitly into the definitions, and although the first variation can be
made into a single vector $\wec{Z}=\sum_{\mu} u_1^{\mu}\pd_{\mu}$, the same
cannot be done for the higher variations. It follows from the fact that a
single field $\wec{Z}$ is enough to define the diffeomorphism $h_l$ which gives
the full transformation of one trajectory onto another, for a small but finite
separation $l$. In practise however, we only know the values of $\wec{Z}$ on
the particular trajectory $\gamma$, and not its dependence on the coordinates
$u$, and we cannot construct the derivatives of $\wec{Z}$ (except in the
$\wec{V}$ direction). That is why the Taylor series of \eqref{Taylor} is
analysed term by term, each with its own equation.
Note that a HVE of a given order $n$ requires the knowledge of the solutions
of all the lower HVE's up to $n-1$, which makes the equations non-homogeneous
(and non-autonomous).

\section{First integrals}

With the present notation, a first integral of the system is a function
$J:M\rightarrow\mathbb{R}$ such that
\begin{equation}
\frac{\de}{\de s}J = \wec{V}(J) = \de J(\wec{V}) = 0.
\end{equation}
A key fact regarding these quantities is that, if there exists a first
integral $J$ of the original system then there also exist first integrals $J_n$
of the variational equations of all orders 
\begin{equation}
    J_n:=\wec{Z}^{(n)}(J).
\end{equation}
To see that these are in fact first integrals of the VE, we make use of the
commutation property again
\begin{equation}
    \frac{\de}{\de s}J_n = \wec{V}(\wec{Z}^{(n)}(J_n)) =
    \wec{Z}^{(n)}(\wec{V}(J)) = 0.
\end{equation}
Just as the HVE, the integral explicitly involves the solutions of the HVE's of
lower order.

However, it can so happen, when taking a particular trajectory $\gamma(s)$ of
$\wec{V}$ along which the variations are considered, that
\[\de J|_{\gamma(s)}\equiv0\]
or in general that
\begin{equation}
    \pd^j J|_{\gamma(s)}\equiv0,\quad \mathrm{for}\quad j<m,
\end{equation}
where $m$ must necessarily be finite or the whole first integral would be a
constant function, and is taken to be the smallest integer with the above
property. Note first that $\pd^m J$ is a well defined tensor of type
$\binom{0}{m}$ because all the partial derivatives of orders lower than $m$
vanish so that only the highest derivative is left
\begin{equation}
    (\pd^m J)_{\gamma(s)}(\underbrace{\wec{Z},\ldots,\wec{Z}}_{m})
    = \underbrace{\wec{Z}(\ldots\wec{Z}}_{m}(J)\ldots)|_{\gamma(s)}.
\label{junior}
\end{equation}
The reason to choose $\pd$ to denote this derivative is that $\nabla$ is
understood to act on the (possible exterior) tangent bundle, while for $\de$
one has $\de^2=0$. It is to be remembered, though, that we do not associate
$\pd$ with derivation in any particular coordinate system, as can be seen in
(\ref{junior}).

Let us take now a VE of $n$-th order with $\pd^n J\equiv0$, and a following
function of the $n$-th variation
\begin{equation}
    \wec{Z}^{(n)}\wec{Z}^{(n')}J = \wec{Z}^{(n)}J_{n'}
\end{equation}
where $n'$ is such that $n+n'=m$ constructed as above. Then
$(\pd^n J_{n'})_{\gamma(s)}$ is a first integral, expressed explicitly as a
function of the $n$-th variation.

Imagine now, that the system has $N$ functionally independent first integrals
in a neighbourhood of $\gamma(s)$. This means that there are $N$ independent
first integrals of the first variational equation (insert Ziglin Lemma), and if
the original integrals are independent on $\gamma(s)$ itself, this means they
all have non-vanishing associated 1-forms $\de J_i$. Consequently we have $N$
independent vectors
\begin{equation}
    \wec{Y}_i = \sharp(\de J_i),
\end{equation}
with which the integrals of the variational equation can be written as
\begin{equation}
    J_{i,1} = \wec{Z}(J_i) = \de J_i(\wec{Z}) = \gm{\wec{Y}_i}{\wec{Z}}.
\end{equation}
Since that expression is constant
\begin{equation}
\begin{aligned}
    0 &= \gm{\nabla_{\wec{V}}\wec{Y}_i}{\wec{Z}} +
    \gm{\wec{Y}_i}{\nabla_{\wec{V}}\wec{Z}}\\
    &= \gm{\dot{\wec{Y}}_i}{\wec{Z}}+
    \gm{\wec{Y}_i}{\nabla_{\wec{Z}}\wec{V}}\\
    &= \gm{\dot{\wec{Y}}_i}{\wec{Z}}+
    \gm{(\nabla\wec{V})^{\dag}\wec{Y}_i}{\wec{Z}},
\end{aligned}
\end{equation}
or
\begin{equation}
    \frac{D}{\pd s}\wec{Y}_i = -
    (\nabla\wec{V})^{\dag}\wec{Y}_i,
    \label{adjoint_y}
\end{equation}
and we say that $\wec{Y}_i$ satisfied the adjoint equation to the variational
equation.
In fact, we have just shown that $N$ first integrals give us a basis
of solutions of the adjoint equation. From them, the solution of the VE can be
obtained by means of the fundamental matrix (or rather operator).

Let $F$ be the fundamental
operator of the VE and the adjoint equation respectively, defined as follows
\begin{equation}
    F = \sum_{i=1}^{N}\wec{Z}_{i0}^{*}\otimes \wec{Z}_i,
\end{equation}
where $\wec{Z}_{i0}$ are the initial conditions of a basis of solutions
$\{\wec{Z}_i\}$, and $\{\wec{Z}^*_i\}$ is the dual basis. The above operator acts
on a constant vector to yield a solution of the variational equation with the
constant vector as the initial value. When the constant vector is prolonged
(along the trajectory $\gamma$)
with the equation $\nabla_{\wec{V}}\wec{Z}_{i0}=0$ (which is not the same as
the VE), then the derivative of the above operator is simply
\begin{equation}
    \frac{D}{\pd s}F = \mathcal{A}F.
\end{equation}
Another way of making the initial values global is to require that the field
$\{\wec{Z}_i\}$ is, at some point, equal to a basis $\{\wec{E}_i\}$ which is
globally parallel: $\nabla\wec{E}_i=0$. This is the usual case with the
impilcit assumption of the base manifold (and connection) being Euclidean.

Define now $P$ to be the fundamental operator of the adjoint equation, and 
$\wec{Y}_{i0}$ be the initial conditions of the appropriate
vector fields; then
\begin{equation}
\begin{aligned}
    c_{ij} &= \gm{\wec{Y}_i}{\wec{Z}_j} = 
    \gm{F\wec{Y}_{i0}}{P\wec{Z}_{j0}}\\ &=
    \gm{P^{\dag}F\wec{Y}_{i0}}{\wec{Z}_{j0}} =
    \gm{\wec{Y}_{i0}}{\wec{Z}_{j0}},
\end{aligned}
\end{equation}
where the last equality is the consequence of this scalar product being
conserved as the first integral. As the initial conditions are arbitrary, this
means that
\begin{equation}
    P = \left(F^{\dag}\right)^{-1}.
\end{equation}

\section{Normal variational equations}

As was mentioned in the chapter on algebraic theory, the variational equation
can be reduced in order by 2, when a first integral is known. We have already
seen how the first part of this reduction works -- by projecting the
variational equation on the subspace orthogonal to the trajectory (tangent
vector). The second step is carried out almost identically, only this time the
vector used for projecting is
\begin{equation}
    \wec{Y}=\sharp\de J,
\end{equation}
where $J$ is the known constant of motion. By definition $\wec{Y}$ is
orthogonal to $\wec{V}$
\begin{equation}
    \gm{\wec{Y}}{\wec{V}}=\de J(\wec{V}) = 0.\label{vy_ort}
\end{equation}
Thus $\prp\wec{Z}$ can be further decomposed as
\begin{equation}
    \prp\wec{Z} = \frac{\gm{\wec{Y}}{\prp\wec{Z}}}{\gm{\wec{Y}}{\wec{Y}}}\wec{Y}
    +\pprp\wec{Z}. \label{2decomp}
\end{equation}
The last vector $\pprp\wec{Z}$ is tangent to the hyper-surface of constant $J$
and orthogonal to $\wec{V}$. Just as $\wec{Z}(J)$ is the first integral of
variational equations $\prp\wec{Z}(J)$ is a first integral of the projected
equations because \eqref{vy_ort} implies
\begin{equation}
    \gm{\wec{Y}}{\wec{Z}} = \gm{\wec{Y}}{\prp\wec{Z}}.
\end{equation}

The reduced variational equation can be obtained from \eqref{2decomp} by
taking the covariant derivative $\nabla_{\wec{V}}$ and projecting on the
subspace orthogonal to both $\wec{V}$ and $\wec{Y}$ or, which amounts to the
same, by applying the Fermi derivative
\begin{equation}
\begin{aligned}
    \frac{D_F}{\pd s}\pprp\wec{Z} &= \frac{D_F}{\pd s}\prp\wec{Z} - 
    \frac{D_F}{\pd s}\left(\frac{\gm{\wec{Y}}{\prp\wec{Z}}}
    {\gm{\wec{Y}}{\wec{Y}}}\wec{Y}\right)\\
    &= \prp\mathcal{A}(\prp\wec{Z}) - \frac{\gm{\wec{Y}}{\prp\wec{Z}}}
    {\gm{\wec{Y}}{\wec{Y}}^2}\left(
    2\gm{\prp\mathcal{A}^{\dag}(\wec{Y})}{\wec{Y}}\wec{Y}
    -\gm{\wec{Y}}{\wec{Y}}\prp\mathcal{A}^{\dag}(\wec{Y})\right),
\end{aligned}
\end{equation}
and projecting with respect to $\wec{Y}$
\begin{equation}
    \pprp\frac{D}{\pd s}\pprp\wec{Z} = \pprp\mathcal{A}\left(\pprp\wec{Z}+
    \frac{\gm{\wec{Y}}{\prp\wec{Z}}}{\gm{\wec{Y}}{\wec{Y}}}\wec{Y}\right)+
    \pprp\mathcal{A}^{\dag}(\wec{Y})
    \frac{\gm{\wec{Y}}{\prp\wec{Z}}}{\gm{\wec{Y}}{\wec{Y}}},
\end{equation}
where the Fermi derivative of $\wec{Y}$ is known from \eqref{adjoint_y}, so
that finally one gets a familiar looking equation
\begin{equation}
    \pprp\frac{D}{\pd s}\pprp\wec{Z} = \pprp\mathcal{A}(\pprp\wec{Z})
    +2\pprp\theta(\wec{Y})
    \frac{\gm{\wec{Y}}{\prp\wec{Z}}}{\gm{\wec{Y}}{\wec{Y}}}.
\end{equation}
In contrast with the (once) projected equation, the above contains an
additional term, which means it is coupled with the degree of freedom parallel
to $\wec{Y}$, albeit the ``coupling'' $\gm{\wec{Y}}{\prp\wec{Z}}$ is constant,
as it is the first integral.

When it comes to the normal variational equation, one is interested in the full
derivative of the variations, not only its transverse value, so that it is
necessary to go back to equation \eqref{full_proj} and change it to
\begin{equation}
\begin{aligned}
    \frac{D}{\pd s}\prp\wec{Z} &= \mathcal{A}(\prp\wec{Z})
    -\mathcal{N}\wec{V}(\gm{\wec{V}}{\wec{Z}})\wec{V}\\
    &=\mathcal{A}(\prp\wec{Z})-
    \mathcal{N}
    \left(\gm{\dot{\wec{V}}}{\wec{Z}}+
    \gm{\wec{V}}{\nabla_{\wec{Z}}\wec{V}}\right)\wec{V}\\
    &=\mathcal{A}(\prp\wec{Z})-
    \mathcal{N}\gm{\dot{\wec{V}}}{\prp\wec{Z}}\wec{V}.
    \label{proj_VE}
\end{aligned}
\end{equation}
The normal part is by definition the variation tangent to the hyper-surface of
the first integral, so that $\gm{\wec{Y}}{\prp\wec{Z}}=0$, and one can simply
write the normal variational equation as
\begin{equation}
    \frac{D}{\pd s}\pprp\wec{Z} = \mathcal{A}(\pprp\wec{Z})
    -\mathcal{N}\gm{\dot{\wec{V}}}{\pprp\wec{Z}}\wec{V}.
    \label{NVE}
\end{equation}

\section{Lyapunov exponents}
Define new operators $L$ and $\Lambda$
\begin{equation}
    L = F^{\dag}F = e^{2(s+1)\Lambda(s)}, \label{L_def}
\end{equation}
so that
\begin{equation}
    \Lambda = \frac{\log(L)}{2(s+1)} = \frac{\log\sqrt{F^{\dag}F}}{s+1}.
\end{equation}
The Lyapunov exponents are then defined as the eigenvalues of
\begin{equation}
    \Lambda_0 = \lim_{s\rightarrow\infty}\Lambda.
\end{equation}
By construction $L$ is self-adjoint, and has positive eigenvalues, so that
$\Lambda$ is well defined. Clearly, the operator $\Lambda_0$ depends on the
particular
solution one uses for $s$ tending to infinity. Or, in terms of $F$, it depends
on the points in the phase space of the variational flow. There is also a
slight difference from the usual notation, where $s$ is present instead of
$s+1$. This change is introduced to make $\Lambda$ well defined for $s=0$ by
formula \eqref{L_def}.

The above formulae require the knowledge of
the fundamental matrix $F$ in order to be able to determine the values of the
exponents. In practise one uses various algorithms \cite{Wolf,Strelcyn} to
reconstruct the
spectrum, and they mostly rely on integration of the variational equation and,
if the system is nonlinear, the original equation as well.

The following considerations provide a differential equation for the Lyapunov
spectrum, or, strictly speaking, an equation determining a operator whose
eigenvalues are the same as those of $\Lambda$ and accordingly tend to the
Lyapunov exponents with $s\rightarrow\infty$. This is achieved through a
similarity transformation
\begin{equation}
    FLF^{-1} = FF^{\dag} = e^{2(s+1)F\Lambda F^{-1}} =: e^{2(s+1)\mathcal L}.
\end{equation}

Let us turn next to the formula for the derivative of matrix (or operator)
exponential
\begin{equation}
    e^{-M}\frac{\de}{\de s}e^M 
    = \int_0^1e^{-\alpha M}\dot{M}e^{\alpha M}\de\alpha. \label{exp_der}
\end{equation}
It is directly applicable to the covariant derivative (or Fermi derivative)
thanks to formula \eqref{op_der}, which could be rewritten as
\begin{equation}
    \frac{D_FM}{\pd s} = \left[\frac{D_F}{\pd s},M\right],
\end{equation}
with each side of the above understood as acting on some vector. The standard
derivation now gives
\begin{equation}
    \frac{\de}{\de\alpha}\left(e^{-\alpha M}\frac{D_F}{\de s}e^{\alpha
    M}\right) =
    e^{-\alpha M}\left[\frac{D_F}{\pd s},M\right]e^{\alpha M} =
    e^{-\alpha M}\frac{D_F M}{\pd s}e^{\alpha M},
\end{equation}
which, when integrated over $\alpha$ leads to \eqref{exp_der}.

Applying the above for $M=2(s+1)\mathcal L$
\begin{equation}
\begin{aligned}
    (FF^{\dag})^{-1}(\dot{F}F^{\dag}+F\dot{F}^{\dag})
    &= \int_0^1e^{-2\alpha(s+1)\mathcal L}
    (2\mathcal L+2(s+1)\dot{\mathcal L})e^{2\alpha(s+1)\mathcal L}\de\alpha,\\
    F^{-\dag}F^{-1}(\mathcal{A}FF^{\dag}+FF^{\dag}\mathcal A^{\dag})
    &= 2\mathcal L +
    2\int_0^1e^{-2\alpha(s+1)\mathcal L}(s+1)\dot{\mathcal L}
    e^{2\alpha(s+1)\Lambda}\de\alpha,\\
    e^{-2(s+1)\mathrm{Ad}\mathcal L}\mathcal A + \mathcal A^{\dag}
    &= 2\mathcal L +2(s+1)\int_0^1 e^{-2\alpha(s+1)\mathrm{Ad}\mathcal L}
    \dot{\mathcal L}\de\alpha,
\end{aligned}
\end{equation}
where the {\it adjoint} (not to be confused with the other adjoint, which
will be explicitly denoted as ``Hermitian adjoint'' throughout this section)
operator has been introduced as
\begin{equation}
    (\mathrm{Ad}W)(U):=[W,U]
\end{equation}
and satisfies
\begin{equation}
    e^{W}Ue^{-W} = e^{\mathrm{Ad}W}U.
\end{equation}
The integral can now be evaluated symbolically (treating the exponent as a
scalar), which is justified by the
direct computation of the exponential series. It gives
\begin{equation}
\begin{aligned}
    (e^{-2(s+1)\mathrm{Ad}\mathcal L}\mathcal - \mathbbm{1})\mathcal{A} + 
    \mathcal A + \mathcal A^{\dag}
    &= 2\mathcal L -\frac{e^{-2(s+1)\mathrm{Ad}\mathcal L}-\mathbbm{1}}
    {\mathrm{Ad}\mathcal L}\dot{\mathcal L},\\
    [\mathcal L,\mathcal A] + \frac{2\mathrm{Ad}\mathcal
    L}{e^{-2(s+1)\mathrm{Ad}\mathcal L}-\mathbbm{1}}\theta 
    &= -\frac{1}{s+1}\mathcal L - \dot{\mathcal L},
\end{aligned}
\end{equation}
or, upon defining
\begin{equation}
    \psi(x) = \frac{x}{1-e^{-x}},
\end{equation}
we finally obtain
\begin{equation}
    \dot{\mathcal L} = \frac{1}{s+1}
    \psi\big(2(s+1)\mathrm{Ad}\mathcal L\big)(\theta)+
    [\mathcal A,\mathcal L] - \frac{1}{s+1}\mathcal L.
    \label{THE}
\end{equation}
Equivalently, by including the $(s+1)$ factor into the operator
$\widetilde{\mathcal L}=(s+1)\mathcal L$, we have
\begin{equation}
    \dot{\widetilde{\mathcal L}} = 
    \psi\big(2\mathrm{Ad}\widetilde{\mathcal L}\big)(\theta)+
    [\mathcal A,\widetilde{\mathcal L}].
\end{equation}
This equation appears to be much simpler than the preceding one, but one has to
remember that $\widetilde{\mathcal L}$ diverges at infinity for non-zero
Lyapunov exponents.

The above derivation requires a few crucial remarks. First of all, the adjoint
operator, trivially has a zero eigenvalue, as any operator commutes with
itself. $\mathrm{Ad}\mathcal L$ is thus not invertible. Thus, it is important
to remember that the fraction notation is meant as symbolic - the ``division''
by an operator is used to write the infinite series obtained by integration as
a simple function. On the other hand, the behaviour of the fraction at ``zero''
(a non-invertible operator) is regular because the function $1/\psi(x)$ has a
removable singularity at $x=0$ and is taken to be equal to its limit there:
$1/\psi(0)=\psi(0)=1$. This also makes clear, that neither $\psi$ nor $1/\psi$
is zero on the real axis.

Although $1/\psi$, when viewed as a series, has infinite radius of
convergence, the same is not true for the other fraction, which is represented
by $\psi$ (the radius at zero is $2\pi$). The function is, however, well
defined for all real values of the argument because its singularities lie on
the imaginary axis. That is why one first replaces the well-behaved series by
$1/\psi$ and then consequently uses only $\psi$.

Finally, the adjoint operator is also Hermitian as $\mathcal L$ is, by
definition, Hermitian. To see this, let us define a natural metric in the vector
space of operators
\begin{equation}
    \gm{A}{B}=\mathrm{tr}(A^{\dag}B),
\end{equation}
so that
\begin{equation}
\begin{aligned}
    \mathrm{tr}(A^{\dag}\mathrm{Ad}\mathcal LB) 
    &= \mathrm{tr}(A^{\dag}\mathcal L B - A^{\dag}B\mathcal L)\\
    &= \mathrm{tr}((\mathcal L A)^{\dag}B-(A\mathcal L)^{\dag}B)\\
    &= \mathrm{tr}([\mathcal L,A]^{\dag}B)\\
    &= \gm{\mathrm{Ad}\mathcal L A}{B},
\end{aligned}
\end{equation}
which means that $\mathrm{Ad}\mathcal L$ has real eigenvalues and $\psi$ is
well behaved on its spectrum.

Equation \eqref{THE} is thus a matrix differential equation with the initial
condition $\mathcal L(0) = \mathbbm{1}$, and involving an operator on the space
of operators $\psi(\mathrm{Ad}\mathcal L)$. Since Ad is linear, the adjoint can
be considered as an $N^2\times N^2$ matrix acting on an $N^2\times1$ vector
(representing a $N\times N$ matrix). The equation is solved for the derivative
which means it is easily implementable numerically. It only requires the
knowledge of a particular solution $\gamma$ around which the linear
approximation is considered. And although the operator $\mathcal L$ is defined
with the use of the fundamental matrix, there is no need of obtaining the basis
of linear solutions to solve for $\Lambda$ because their spectra are the same.

The dimension of the equation seems to complicate matters a lot
because, for example, a Hamiltonian system of two spatial degrees of freedom,
which requires a four dimensional phase space, gives rise to a $16\times16$
adjoint matrix. Evaluating $\psi$ on such a matrix cannot be achieved
by a series, as mentioned earlier, and requires an eigenvalue decomposition,
which would make the calculations cumbersome. Fortunately, this is not a
general $N^2\times N^2$ operator, and the knowledge of $\mathcal L$ is all we
need. 

Take any Hermitian operator $A$ with an orthonormal basis of eigenvectors
$\{\wec{U}_i\}$ such that $A(\wec{U}_i)=\lambda_i \wec{U}_i$, another operator
$B$, constructed from those eigenvectors, and forms of the dual basis
\begin{equation}
    B= \wec{U}^*_k\otimes \wec{U}_l,
\end{equation}
for given $k$ and $l$. The action of the adjoint of $A$ on $B$ is as follows
\begin{equation}
\begin{aligned}
    AB &= \wec{U}^*_k\otimes A(\wec{U}_l)=
    \lambda_l \wec{U}^*_k\otimes \wec{U}_l,\\
    (BA)^{\dag} &= AB^{\dag} = \wec{U}^*_l\otimes A(\wec{U}_k) = \lambda_k
    \wec{U}^*_l\otimes
    \wec{U}_k,
\end{aligned}
\end{equation}
so that $BA=\lambda_k \wec{U}^*_k\otimes \wec{U}_l$ and
\begin{equation}
    \mathrm{Ad}A(B) = (\lambda_l-\lambda_k)B,
\end{equation}
where the previously obtained properties of Hermitian adjoint of a simple
tensor product of orthonormal bases were used.

We thus have constructed a full set of eigenvectors and eigenvalues, of which
$N$ are identically zero. This knowledge makes the practical computation of
$\psi(\mathrm{Ad}\mathcal L)$ much faster and, in theory, allows of writing
all terms of equation \eqref{THE} explicitly when the characteristic
polynomial is soluble.

\section{On some additional properties}

This section is devoted to describing how the Lyapunov exponents, and the
system in general, behave when there are special constraints
present. The first is simply the Hamiltonian structure, and the second is the
more general conserved integral invariant of the flow.

The Hamiltonian structure is usually introduced by means of a symplectic form
$\omega_S$,
but here, since there is already a distinguished metric structure, a new
operator $\mathcal{I}$ ($\binom{1}{1}$ tensor) can be used to create symplectic
structure. We have
\begin{equation}
    \omega_S(\wec{X},\wec{U}) = \gm{\mathcal{I}\wec{X}}{\wec{U}},
\end{equation}
so that for the Hamiltonian $H$ the associated vector field is
\begin{equation}
    \wec{V} = \mathcal{I}(\sharp\de H) \iff 
    \omega_S(\wec{V},\wec{U}) = \de H (\wec{U}).
\end{equation}
This will also mean that there is a particular coordinate system
$\{q,p\}$ for which the coordinate basis is orthonormal and, consequently, in
which the connection is Euclidean, i.e.
$\nabla\pd_q=\nabla\pd_p=0$, so that
\begin{equation}
    \mathcal{I} =
    \sum_{i=1}^{K}(\de{p_i}\otimes\pd_{q_i}-\de{q_i}\otimes\pd_{p_i}),
\end{equation}
where as before $2K=N$ is the dimension of the manifold. By definition the
operator is anti-Hermitian $\mathcal{I}^{\dag}=-\mathcal{I}$, anti-involutive
$\mathcal{I}^2=-\mathbbm{1}$ and, because the connection is euclidean, it is
also parallel $\nabla\mathcal{I}=0$.

As follows from previous sections $\gm{\sharp\de H}{\wec{Z}}=\mathrm{const}$
and $\gm{\sharp\de H}{\wec{V}}=0$. Consider next two solutions of the
variational equation $\wec{Z}_1$ and $\wec{Z}_2$ and the question of
conservation
\begin{equation}
\begin{aligned}
    \wec{V}(\gm{\wec{Z}_1}{\mathcal{I}\wec{Z}_2}) &= 
    \gm{\mathcal{A}\wec{Z}_1}{\mathcal{I}\wec{Z}_2}+
    \gm{\wec{Z}_1}{\mathcal{I}\mathcal{A}\wec{Z}_2}\\
    &=\gm{\mathcal{I}\nabla_{\wec{Z}_1}(\sharp\de H)}{\mathcal{I}\wec{Z}_2}-
    \gm{\wec{Z}_1}{\nabla_{\wec{Z}_2}(\sharp\de H)}\\
    &=\gm{\sharp\de H}{\nabla_{\wec{Z}_1}\wec{Z}_2}-
    \gm{\nabla_{\wec{Z}_2}\wec{Z}_1}{\sharp\de H}\\
    &= \de H([\wec{Z}_1,\wec{Z}_2])\\
    &=\wec{Z}_1(\mathrm{const})-\wec{Z}_2(\mathrm{const})=0,
\end{aligned}
\end{equation}
where the appearance of the commutator follows from the zero torsion condition.
The above product is thus conserved, and this means the following for the
fundamental operator
\begin{equation}
\begin{aligned}
    \gm{F\wec{Z}_{10}}{\mathcal{I}F\wec{Z}_{20}} &=
    \gm{\wec{Z}_{10}}{\mathcal{I}\wec{Z}_{20}},\\
    F^{\dag}\mathcal{I}F &= \mathcal{I},\\
    FF^{\dag} &= -F\mathcal{I}F^{-1}\mathcal{I},\\
    (FF^{\dag})^{-1} &= -\mathcal{I}F\mathcal{I}F^{-1} =
    \mathcal{I}(FF^{\dag})\mathcal{I}^{-1},
\end{aligned}
\end{equation}
so that the operator $L$ defined in the previous section has the same
eigenvalues as its inverse, which means the Lyapunov matrix $\mathcal{L}$ has
pairs of eigenvalues of opposite signs $\{\lambda,-\lambda\}$. This also
implies the flow conserves the phase space volume, as the sum of all such
eigenvalues is zero. It should be stressed, that although the symplectic
structure is enough to define volume as $\eta = \omega_S^{\wedge K}$, it does
not
give a metric structure and the respective Levi-Civitta connection. This is the
reason for the introduction of the particular canonical basis $\{\pd_q,\pd_p\}$ 
defined as orthonormal.

This brings us to the next property, which is preserving the integral of an
$N$-form, only this time it needs not be the volume form $\eta$. Because
the space of $N$-forms is one dimensional the new quantity is a multiple of
$\eta$, say $\alpha\eta$. Let us assume that
\begin{equation}
    \int_{D(s)}\alpha\eta = \int_{D_0} g^{*}_s(\alpha\eta) = \mathrm{const},
\end{equation}
where $D(s)$ is the image of some region $D_0$ through the diffeomorphism
$g_s$. Differentiating one gets
\begin{equation}
\begin{aligned}
    \int_{D_0}\pounds_{\wec{V}}(\alpha\eta) &= 
    \int_{D_0}\de(\iota_{\wec{V}}(\alpha\eta))+
    \iota_{\wec{V}}\de(\alpha\eta)\\
    &= \int_{D_0}\de(\iota_{\alpha\wec{V}}\eta)\\
    &= \int_{D_0}\pounds_{\alpha\wec{V}}\eta\\
    &=: \int_{D_0}\mathrm{div}(\alpha\wec{V})\eta.
\end{aligned}
\end{equation}
The usual notion of divergence-free flows is just a special case when
$\alpha=1$. To see how this definition of divergence works with the volume, and
the how to compute with the covariant derivative, recall first an identity for
the Lie derivative
\begin{equation}
    (\pounds_{\wec{V}}\eta)(\wec{E}_1,\ldots,\wec{E}_N) =
    \wec{V}(\eta(\wec{E}_1,\ldots,\wec{E}_N)) -
    \sum_{i=1}^N \eta(\wec{E}_1,\ldots,[\wec{V},\wec{E}_i],\ldots,\wec{E}_N),
\end{equation}
for any basis $\{\wec{E}_i\}$. Together with the equality \eqref{par_vol}, and
the fact that the torsion is zero, the above amounts to
\begin{equation}
\begin{aligned}
    (\pounds_{\wec{V}}\eta)(\wec{E}_1,\ldots,\wec{E}_N)&=
    \sum_{i=1}^N
    \eta(\wec{E}_1,\ldots,\nabla_{\wec{E}_i}\wec{V},\ldots,\wec{E}_N),\\
    \mathrm{div}(\wec{V})\eta(\wec{E}_1,\ldots,\wec{E}_N)&=
    \sum_{i=1}^N \eta(\wec{E}_1,\ldots,
    \wec{E}^*_i(\nabla_{\wec{E}_i}\wec{V})\wec{E}_i,\ldots,\wec{E}_N),\\
    \mathrm{div}\wec{V} &= \mathrm{tr}(\nabla\wec{V}),
\end{aligned}
\end{equation}
where the middle line is a direct consequence of the volume form being
completely anti-symmetric and $\{\wec{E}_i\}$ constituting a basis. To put it
shortly, a invariant measure exists when there exists what is called the last
multiplier $\alpha$ such that
\begin{equation}
    \mathrm{tr}(\nabla(\alpha\wec{V}))=0.
\end{equation}

The special case of $\alpha=1$ (which holds for Hamiltonian systems, but not
only), means simply that the flow conserves volume, as can be seen from the
integral formulation above. This has a straightforward consequence on the
Lyapunov exponents, since
\begin{equation}
    \eta(\wec{Z}_1,\ldots,\wec{Z}_N)=(\det F)
    \eta(\wec{Z}_{10},\ldots,\wec{Z}_{N0}),
\end{equation}
which holds for any $N$-form. The determinant changes according to
\begin{equation}
\begin{aligned}
    \wec{V}(\eta(\wec{Z}_1,\ldots,\wec{Z}_N)) &= \sum_{i=1}^N
    \eta(\wec{Z}_1,\ldots,\mathcal{A}\wec{Z}_i,\ldots,\wec{Z}_N),\\
    \frac{\de}{\de s}(\det F)\eta(\wec{Z}_{10},\ldots,\wec{Z}_{N0}) &=
    \mathrm{tr}(\mathcal{A})\eta(\wec{Z}_1,\ldots,\wec{Z}_N),\\
    \frac{\de}{\de s}(\det F) &= \mathrm{tr}(\mathcal{A})\det F.
\end{aligned}
\end{equation}
Since $2\mathrm{tr}(\mathcal{L}) = \ln\det(FF^{\dag})= 2\ln|\det F|$, the
exponents add up to zero when the divergence of $\wec{V}$ vanishes.

\chapter{Examples}

The following dynamical systems will be used to show how the Lyapunov exponents
equation \eqref{THE} can be used in practise, and also how the normal
variational equations are computed in the geometric context versus the
algebraic one. Only the geometric considerations are in fact new, as the
Galoisian obstructions to integrability of these systems were all analysed in
details in the papers cited in each respective section.

In practise, the evolution of the exponents is considered in the time $t$ which
is not the natural parameter, so that the results coincide with the standard
ones. This does not change any of the formulae, as the requirement that the
vector field $\wec{V}$ be normalised only matters when the projections are
introduced, while the exponents are calculated for the full $N$ dimensional
system. The equation \eqref{THE} itself is then integrated using the
Runge-Kutta method of the fourth order.

\section{Arnold-Beltrami-Childress flow}

The system is given by
\begin{equation}
    \wec{W}=
    \begin{pmatrix}
    \dot{x}\\
    \dot{y}\\
    \dot{z}
    \end{pmatrix} =
    \begin{pmatrix}
    A \sin z + C\cos y\\
    B \sin x + A\cos z\\
    C \sin y + B\cos x
    \end{pmatrix}.
    \label{ABC_V}
\end{equation}

We consider $A^2=B^2$ and $ABC\ne 0$. By shifting the variables, all parameters
can be made positive, and as was shown in \cite{Maciejka_ABC} the system is
not integrable in the sense that there is no meromorphic first integral on the
complex torus $T_{\mathbb C}^3$ which is the system's phase space and if
$\mu^2:=C^2/(2A^2)\leq 1$ there are no real integrals either. Only one
first integral is needed for integrability, because the flow has also zero
divergence and accordingly has a trivial last multiplier. For a three
dimensional system that is enough to prove there must also exist two additional
integrals \cite{Goriely}. Obviously for $\mu=0$ the system is separable, and
solvable, and this facts allows for testing how the Lyapunov exponents behave.

\begin{figure}[h]
\includegraphics[width=\textwidth]{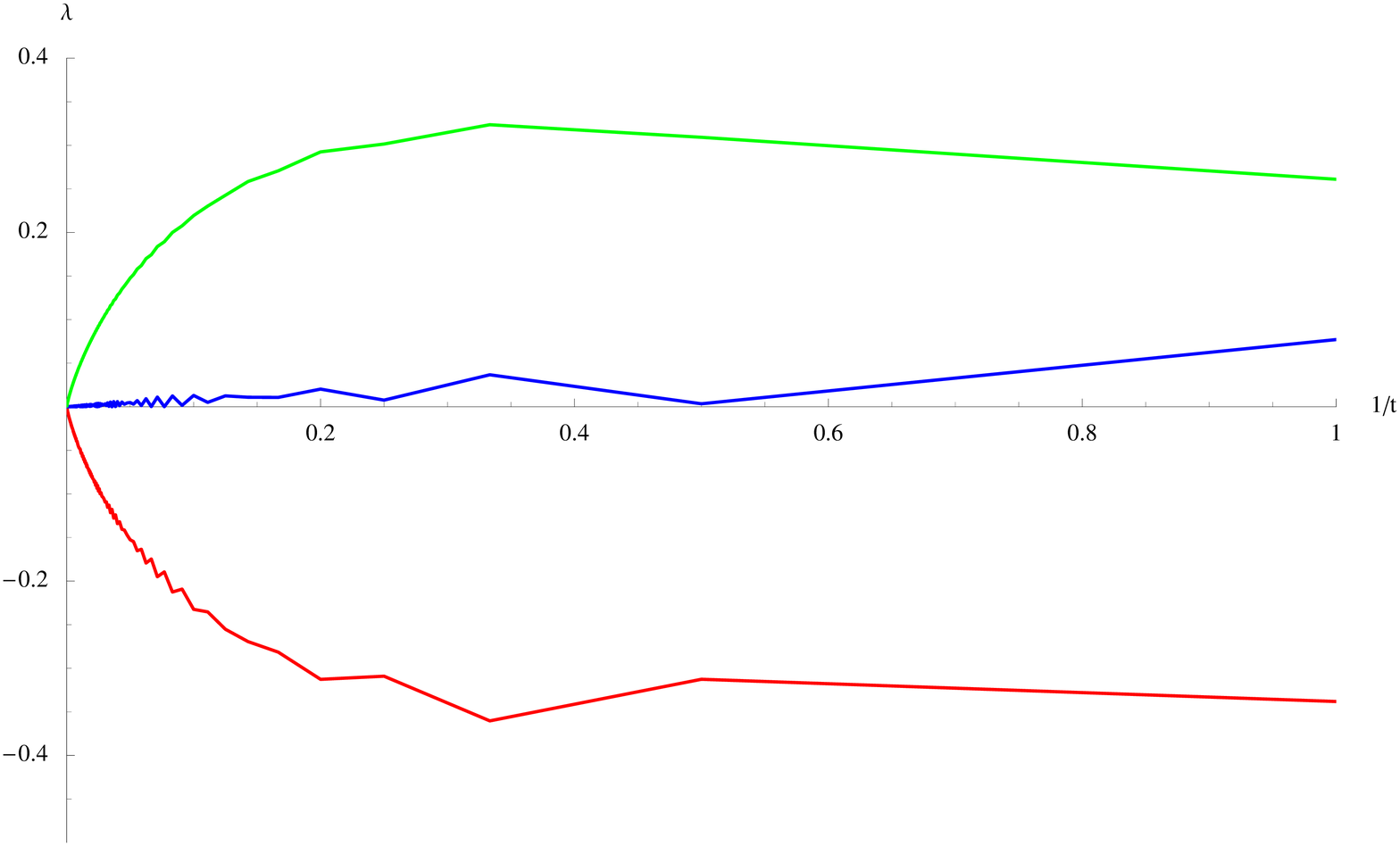}
\caption{The Lyapunov exponents for the ABC flow in the integrable case
$\mu=0$.}
\label{ABC_0}
\end{figure}

\begin{figure}[h]
\includegraphics[width=\textwidth]{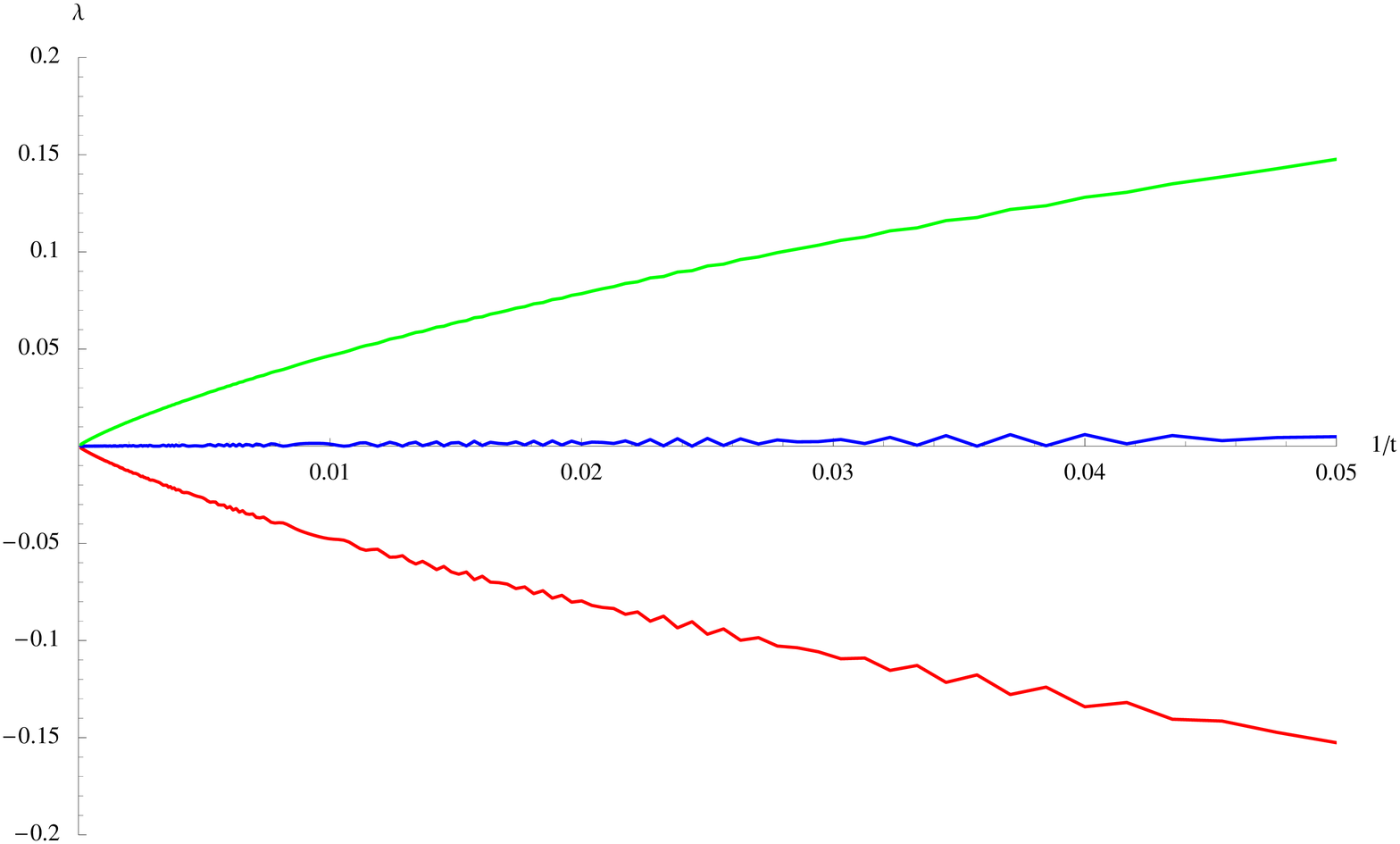}
\caption{The Lyapunov exponents for the ABC flow in the integrable case
($\mu=0$) - a
magnification of the $1/t=0$ vicinity.}
\label{ABC_0a}
\end{figure}

First, the integrable case with initial conditions $x(0)=y(0)=0$ and
$z(0)=0.001$. Using formula \eqref{THE} the exponents are evolved in time $t$,
and can be plotted as functions of $1/t$ -- so that the origin of the
horizontal axis
corresponds to $t\rightarrow\infty$. The results are presented in figures
\ref{ABC_0} and \ref{ABC_0a}.

The maximal value of $t$ was $10000$, and the values of the exponents at that
point are
\[\lambda=(-9.173\times10^{-4},9.099\times10^{-4},7.404\times10^{-7}).\]

\begin{figure}[h]
\includegraphics[width=\textwidth]{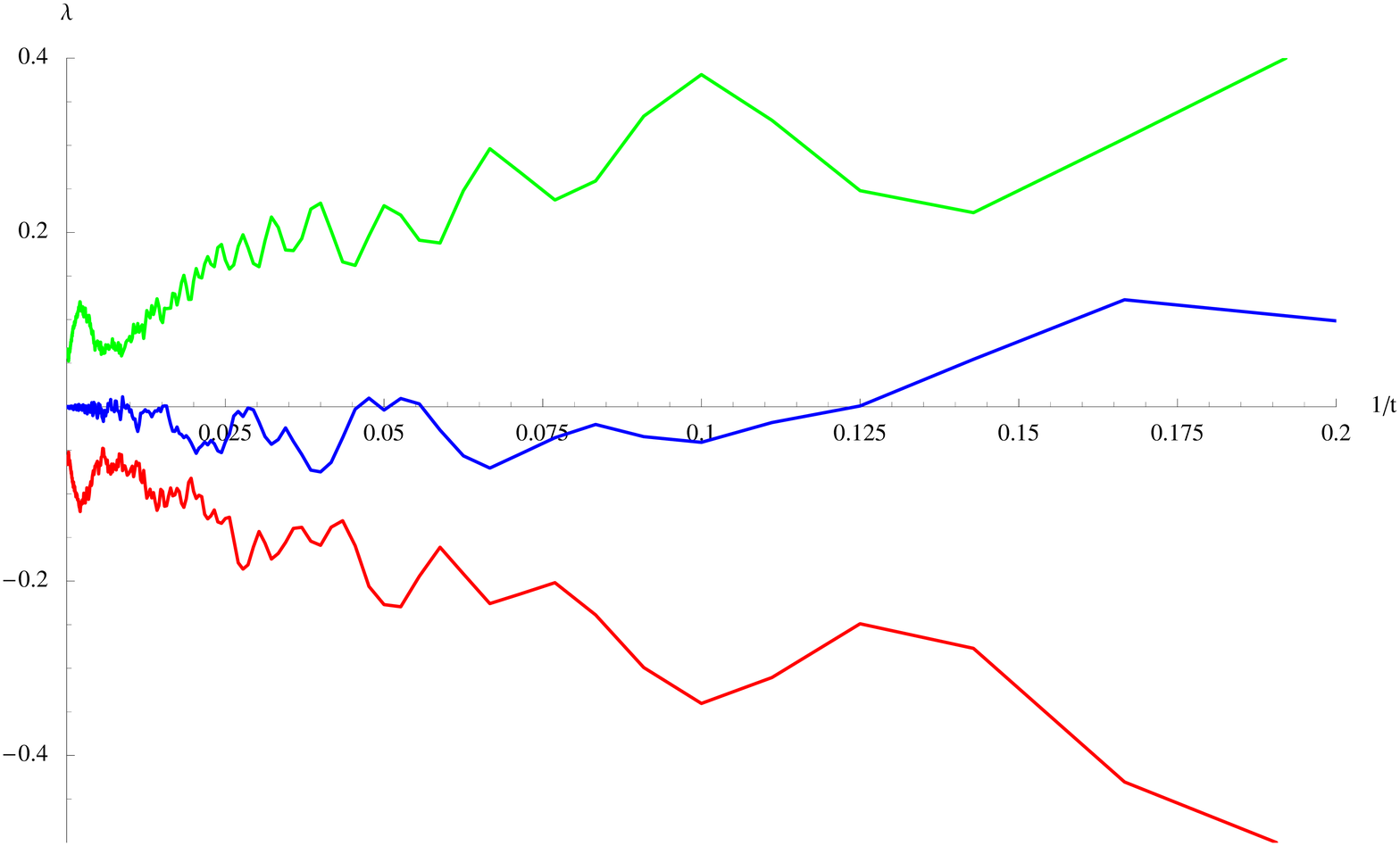}
\caption{The Lyapunov exponents for the ABC flow in a non-integrable case
$\mu=1/2$.}
\label{ABC_1}
\end{figure}

\begin{figure}[h]
\includegraphics[width=\textwidth]{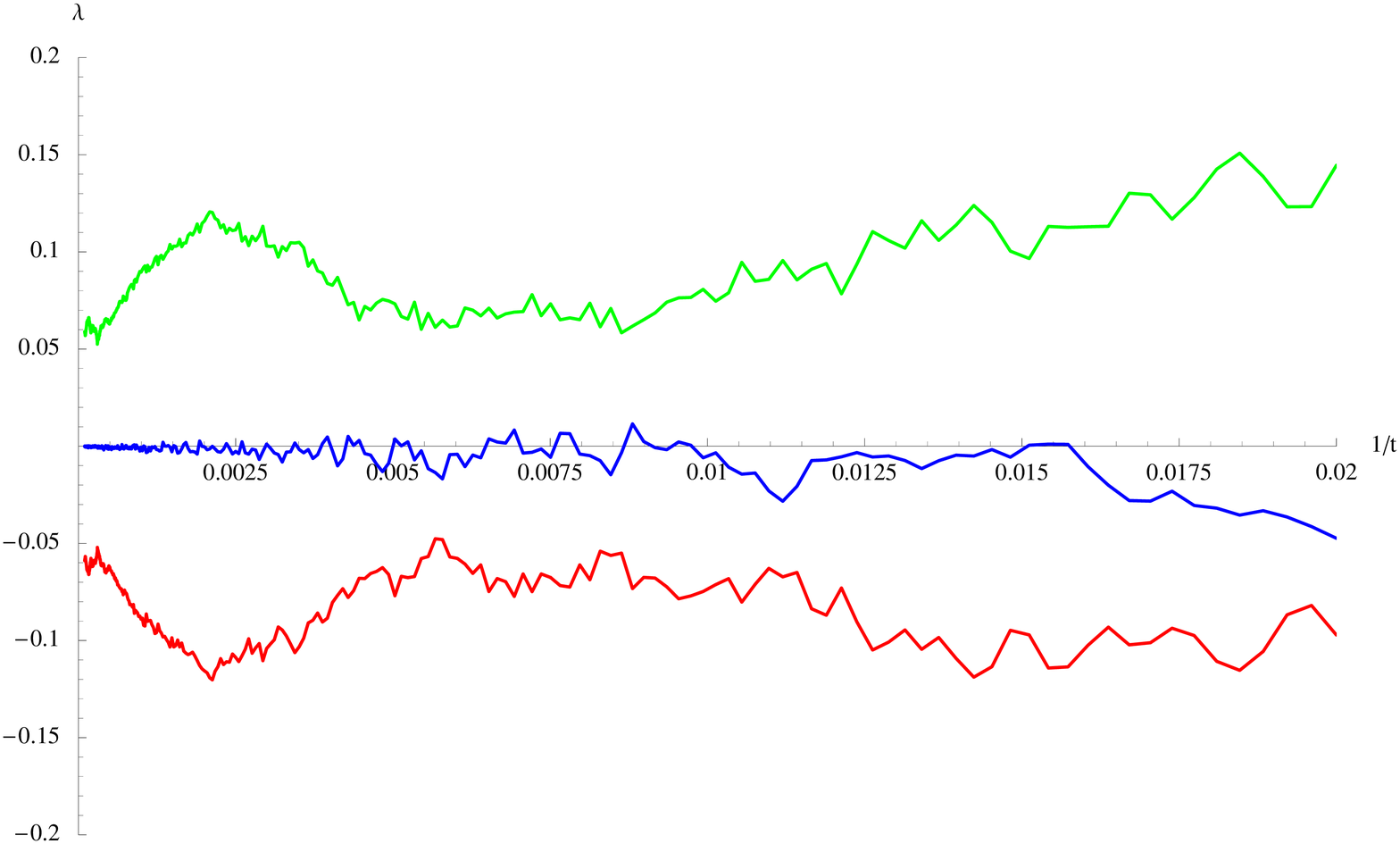}
\caption{The Lyapunov exponents for the ABC flow in a non-integrable case
($\mu=1/2$) - a magnification of the $1/t=0$ vicinity.}
\label{ABC_1a}
\end{figure}

Taking now the value of $\mu=1/2$ to investigate a non-integrable scenario, and
the same initial conditions, the exponents can be seen to no longer all be
zero, there is one tending to zero and two non zero of opposite signs, as it to
be expected for a flow with conserved volume. Their values at $t=10000$ are
\[\lambda=(5.9\times10^{-2},-5.886\times10^{-2},-1.385\times10^{-4}),\]
and their time
dependence is shown in figures \ref{ABC_1} and \ref{ABC_1a}.

When it comes to the variational equation, since there are no known first
integral (a priori), we can only go as far as projecting it with respect to the
trajectory. As shown in \cite{Maciejka_ABC} there is a particular solution for
which $y=\pi/4$, $z=x+\pi/2$, so that the trajectory is described by a single
equation in $x(t)$ and gives as the projected VE
\begin{equation}
\begin{aligned}
    \dot{\xi} &= \xi\sin(x)-2\mu\eta\\
    \dot{\eta} &= \xi\cos(x),
\end{aligned}
\end{equation}
where $\xi$ and $\eta$ are the variations.

In the other approach, taking the coordinate basis $(\pd_x,\pd_y\pd_z)$ to be
orthonormal, equation \eqref{proj_VE} becomes
\begin{equation}
    \frac{\de}{\de s}\prp Z^{\beta} = 
    (\mathcal{A}^{\beta}_{\nu} - V^{\beta}\dot{V_{\nu}}) \prp Z^{\nu}.
\end{equation}
which we want to change to involve time $t$, to compare it with the equations
obtained above. Since the relation between the vectors is
$\wec{W}=\gm{\wec{W}}{\wec{W}}\wec{V}$, the required derivations are rather
cumbersome, but using the particular solution and the relation $\de s =
\sqrt{\gm{\wec{W}}{\wec{W}}}\de t$, the above is finally reduced to
\begin{equation}
    \prp\dot{Z}^{\beta} = 
    \tilde{\mathcal{A}}^{\beta}_{\nu}\prp Z^{\nu},
\end{equation}
with
\begin{equation}
    \tilde{\mathcal{A}} =
    \frac{A}{2} \begin{pmatrix}
    \sin(x) & -2\mu & -\sin(x) \\
    2\cos(x) & 0 & -2\cos(x) \\
    -\sin(x) & 2\mu & \sin(x)
    \end{pmatrix}.
\end{equation}
This is in fact the same equation as before, when one defines $\xi$ to be
\mbox{$\prp Z^x-\prp Z^z$} and $\eta = \prp Z^y$ (in the cited paper the
coefficient $A$ was taken to be 1). Obviously only two degrees
of freedom are needed after the projection.

The geometric thus gives the same ``starting'' equation for further
investigation. Its details -- the determination of the differential Galois
group can be found in the cited article.

\section{$so(5)$ Gross-Neveu system}
The Hamiltonian reads
\begin{equation}
    H = \frac{1}{2}(p_1^2+p_2^2)-2\cos(q_1)-2\cos(q_2)-4\cos(q_1)\cos(q_2),
\end{equation}
which translates into the appropriate vector $\wec{W}$
\begin{equation}
    \frac{\de}{\de t}
    \begin{pmatrix}
    q_1\\
    q_2\\
    p_1\\
    p_2
    \end{pmatrix}
    = 
    \begin{pmatrix}
    p_1\\
    p_2\\
    -2\sin(q_1)-4\sin(q_1)\cos(q_2)\\
    -2\sin(q_2)-4\cos(q_1)\sin(q_2)
    \end{pmatrix}.
\end{equation}

As mentioned before, this system is never integrable meromorphically in the
Liouville sense as shown it \cite{GN}. As the cited paper indicates, it has
the interesting feature of appearing regular in the original variables, and
clearly chaotic when a complex canonical transformation $q\rightarrow i q$,
$p\rightarrow -i p$ is performed. The equations presented above are those
after the transformation. In both cases the ``effective'' coordinates
remain real, that is, when the initial conditions are real the variables
remain real, and when they start as imaginary, they remain purely imaginary.

\begin{figure}[h]
\includegraphics[width=\textwidth]{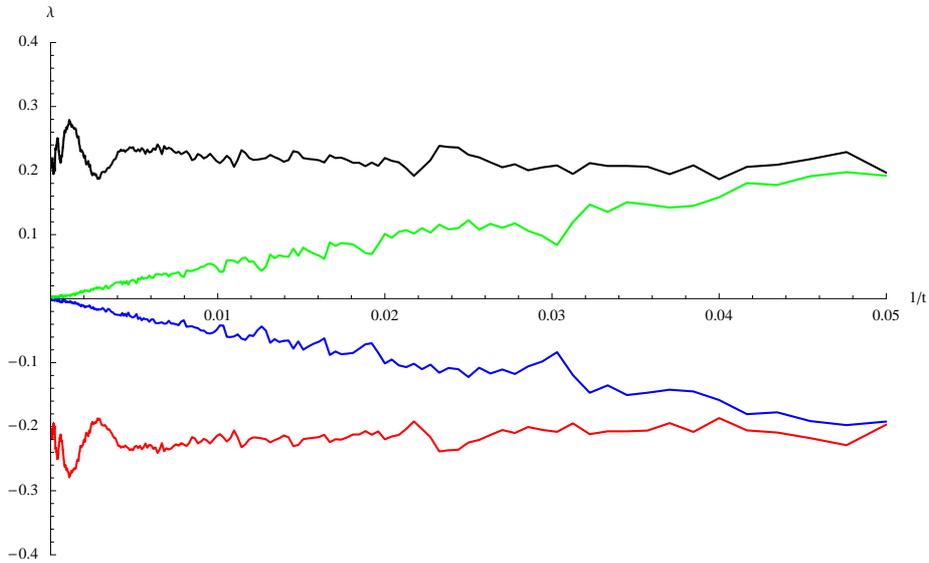}
\caption{The Lyapunov exponents for the Gross-Neveu system in the imaginary
domain.}
\label{GN_0}
\end{figure}

The first set of Lyapunov exponents was obtained for the imaginary domain
(explicitly the above Hamiltonian) for the initial conditions of $q_1(0)=0.01$,
$q_2(0)=0$, $p_1(0)=0.01$ and $p_2(0)$ positive, determined by the condition
$H=3$. At the maximal time of $10000$ the spectrum was
\[\lambda=(0.2169, -0.2169, -2.64\times10^{-3}, 2.64\times10^{-3}).\]
It is clear that
two of the exponents remain non zero as depicted in figure \ref{GN_0}.

\begin{figure}[h]
\includegraphics[width=\textwidth]{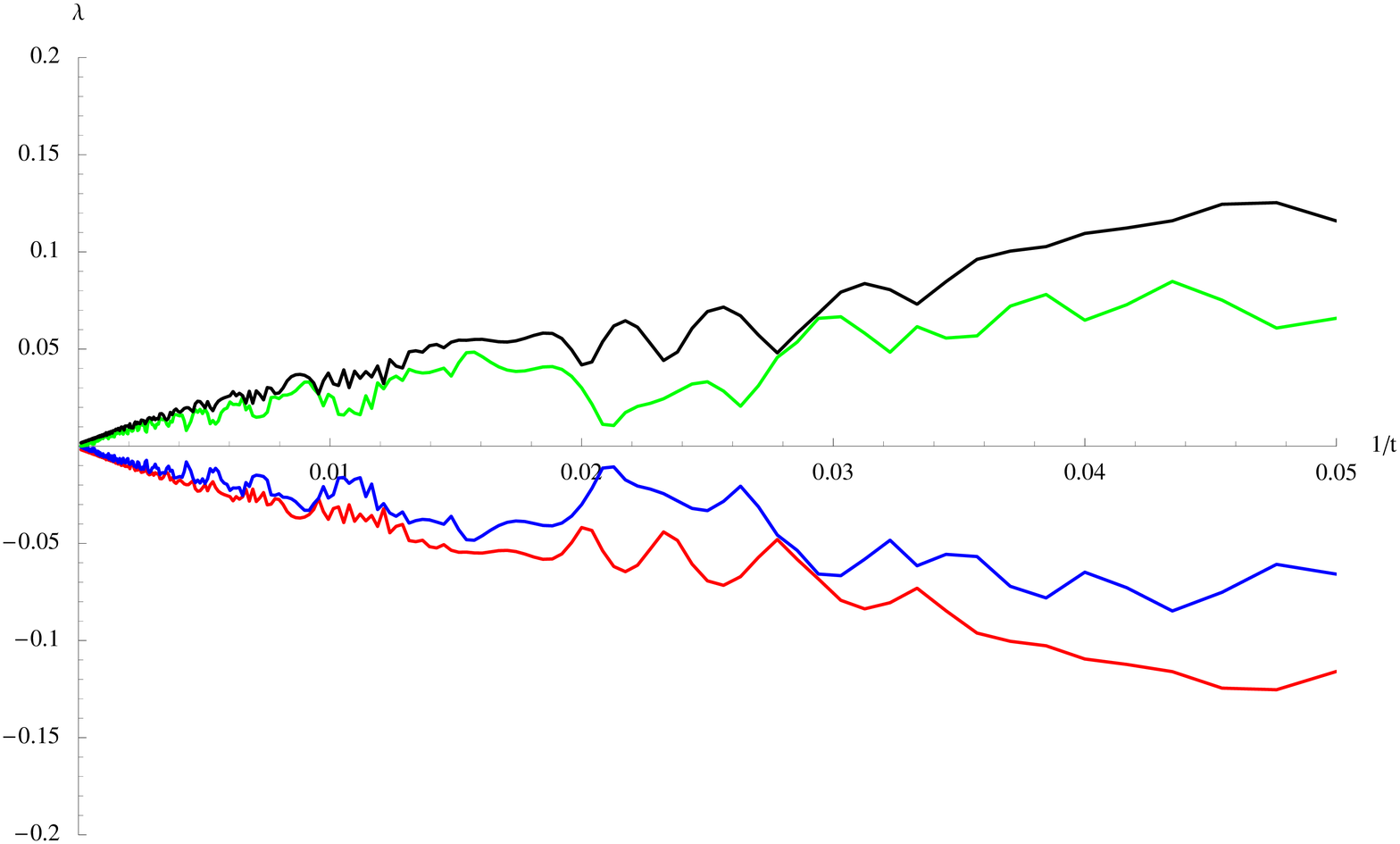}
\caption{The Lyapunov exponents for the Gross-Neveu system in the real
domain.}
\label{GN_I}
\end{figure}

For the real domain, where the Hamiltonian becomes
\begin{equation}
    H = -\frac{1}{2}(p_1^2+p_2^2)
    -2\cosh(q_1)-2\cosh(q_2)-4\cosh(q_1)\cosh(q_2),
\end{equation}
taking the initial conditions of $q_1(0)=0.01$, $q_2(0)=0$, $p_1(0)=0.01$ and
$p_2(0)$ positive such that $H=-3$, the exponents all tend to zero with
\[\lambda=(1.753\times10^{-3}, -1.753\times10^{-3}, -3.651\times10^{-4},
3.651\times10^{-4})\]
at $t=10000$. The results for this case is shown in figure
\ref{GN_I}.

Let us see now how the double projection works to produce the normal
variational equations in this case. In the algebraic approach the invariant
plane $q_2=p_2=0$ is used to find a particular solution $(q_1(t),p_1(t))$, for
which the Jacobian matrix of $\wec{W}$ is
\begin{equation}
    A=\begin{pmatrix}
    0 & 0 & 1 & 0\\
    0 & 0 & 0 & 1\\
    -6\cos(q_1(t)) & 0 & 0 & 0\\
    0 & -2-4 \cos(q_1(t)) & 0 & 0
    \end{pmatrix}.
\end{equation}
Taking now only the variations in the directions $q_2$ and $p_2$ we get the NVE
\begin{equation}
\begin{aligned}
    \dot{\xi} &= \eta\\
    \dot{\eta} &= -2-4\cos(q_1(t)),
\end{aligned}
\end{equation}
where $\xi$ and $\eta$, as before, are the variations.

The geometric procedure is essentially the same as for the ABC flow, in that it
consists of normalising the field $\wec{W}$, and the appropriate equation
\eqref{NVE} is now
\begin{equation}
    \pprp\dot{\wec{Z}} = \begin{pmatrix}
    -36 f \sin^2(q_1) & 0 & -6fp1\sin(q1) & 0\\
    0 & 0 & 0 & 1\\
    -6fp1\sin(q_1) & 0 & -6fp_1^2 & 0 \\
    0 & -2 -4\cos(q_1) & 0 & 0
    \end{pmatrix}\pprp\wec{Z},
\end{equation}
with
\begin{equation}
    f = \frac{6p_1(6\cos(q_1)-1)\sin(q_1)}{(p_1^2+36\sin^2(q_1))^2}.
\end{equation}
This is, again, the same as the algebraic NVE, when two of the degrees of
freedom corresponding to $p_1$ and $q_1$ are suppressed. Alternatively one can
check that the two vectors with respect to which the projection takes place,
span the $p_1$, $q_1$ subspace, because on the trajectory
\begin{equation}
\begin{aligned}
    \wec{W} &= (p1, 0, -6\sin(q_1),0),\\
    \wec{Y} = \sharp\de H &= (6\sin(q_1),0,p1,0).
\end{aligned}
\end{equation}

The above means that the next steps -- checking if the NVE are soluble in the 
Liouvillian sense -- is the same in both approaches. The proof that there are
no such solutions can be found in the paper cited at the beginiing of this
section.

\section{Friedmann-Robertson-Walker cosmology}

The last example is a cosmological system obtained for the FRW universe with a
scalar field conformally coupled to gravity. It was analysed in great detail in
\cite{FRW}, and includes both integrable and non-integrable sub-cases.

The particular Hamiltonian taken here is
\begin{equation}
    H = \frac{1}{2}(p_1^2+p_2^2)-\frac{1}{2}m^2q_1^2q_2^2+
    \frac{1}{4}(\Lambda q_1^4+\lambda q_2^4).
\end{equation}
When
\[\Lambda=\lambda=-m^2\]
there is another first integral
\begin{equation}
    J = q_1p_2-q_2p_1,
\end{equation}
and no additional integral exists if the parameters are varied slightly. It is
thus convenient to substitute $m^2=-\lambda\epsilon$, so that the vector
$\wec{V}$ is
\begin{equation}
    \frac{\de}{\de t}\begin{pmatrix}
    q1\\q2\\p1\\p2\end{pmatrix} =
    \begin{pmatrix}
    p1\\p2\\
    -\lambda q_1(q_1^2+q_2^2(1+\epsilon))\\
    -\lambda q_2(q_2^2+q_1^2(1+\epsilon))
    \end{pmatrix}.
\end{equation}

\begin{figure}[h]
\includegraphics[width=\textwidth]{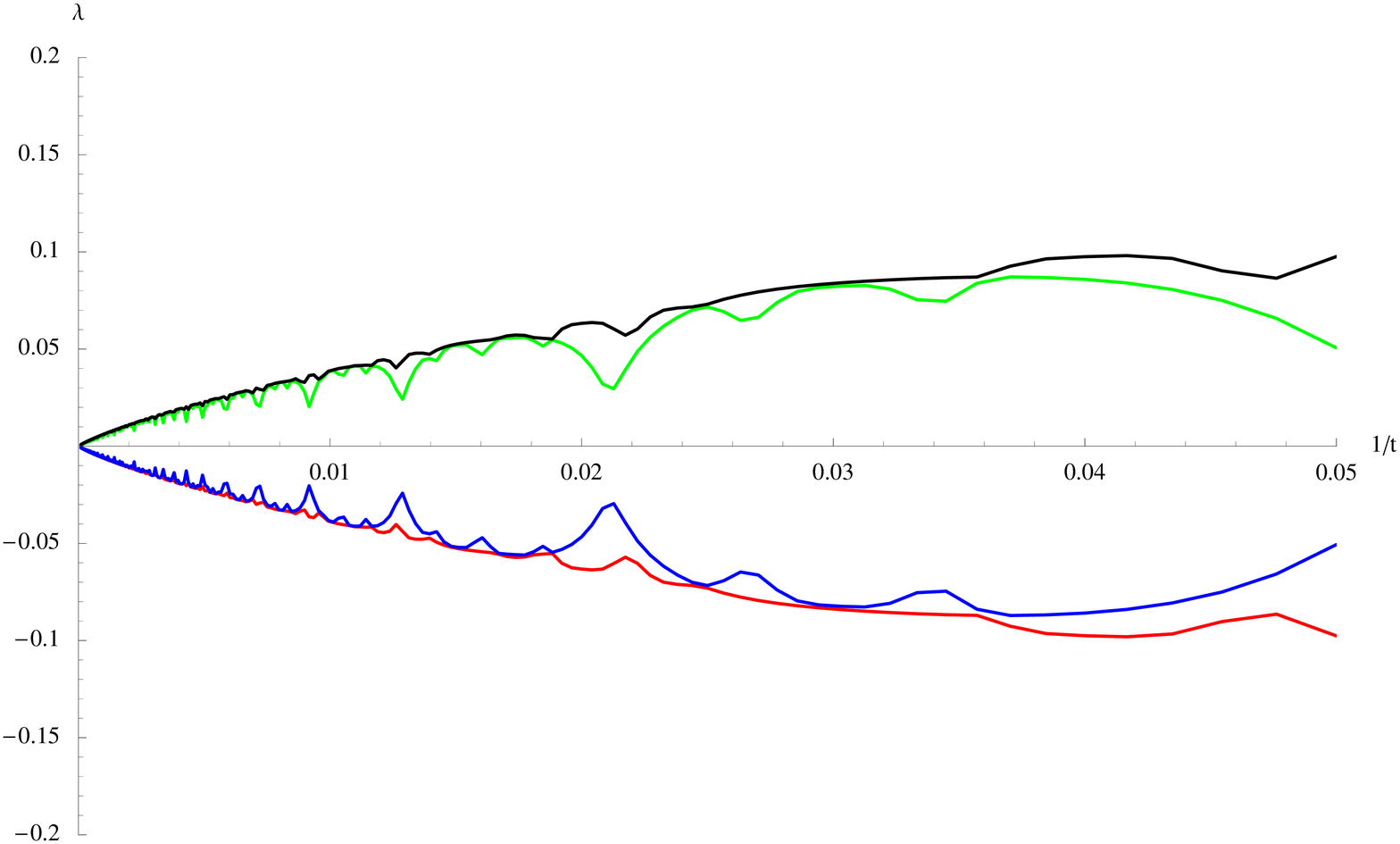}
\caption{The Lyapunov exponents for the FRW system for $\epsilon=-1$.}
\label{FRW_0}
\end{figure}

This example is used to show, that even though the system is integrable only
when $\epsilon=0$, the Lyapunov exponents remain zero until the
``perturbation'' is big enough. Specifically, when $\epsilon=-1$, the spectrum
at $t=10000$ is 
\[\lambda=
(-8.461\times10^{-4}, 8.461\times10^{-4},
-7.909\times10^{-4}, 7.909\times10^{-4}).\]
Its time evolution is presented in figure \ref{FRW_0}.
The initial conditions were $q_1(0)=0.01$, $q_2(0)=0$, $p_1(0)=0.01$ and
$p_2(0)$ positive, determined by $H=0.01$.

Changing the value of $\epsilon$ to $-1.9$, with the same initial conditions
gives a completely different picture. The exponents now read
\[\lambda=(-0.1388, 0.1388, -8.078\times10^{-4}, 8.078\times10^{-4}).\]
This is to be
expected for a Hamiltonian system -- that the chaos becomes visible only for
sufficient perturbation.

\begin{figure}[h]
\includegraphics[width=\textwidth]{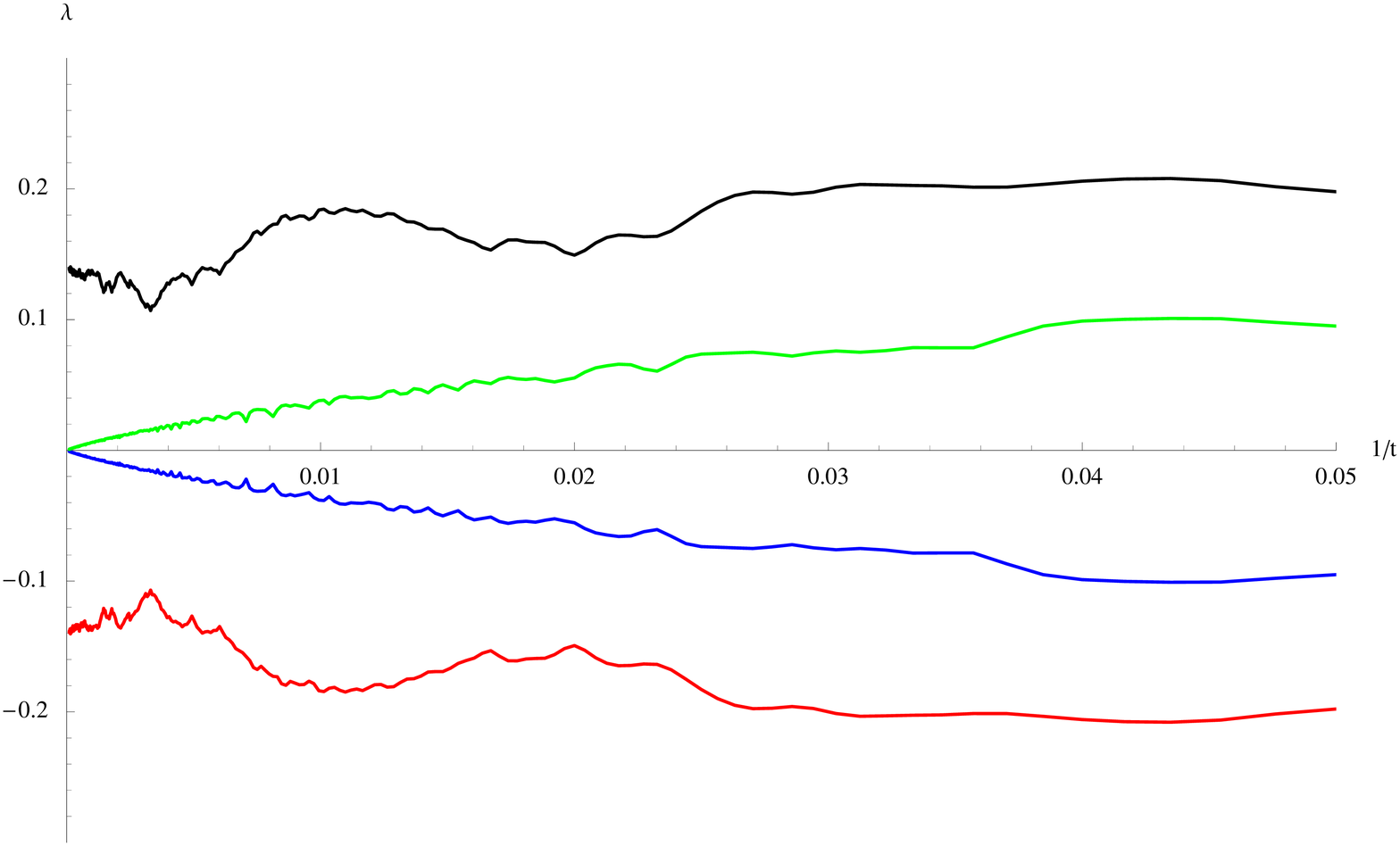}
\caption{The Lyapunov exponents for the FRW system for $\epsilon=-1.9$.}
\label{FRW_1}
\end{figure}

\chapter{Final Matters}

The study presented here can only be considered a beginning of further
exploration of the subject, but it can be definitely said, that the two
approaches -- geometric and algebraic -- can be successfully conflated to yield
a better insight into integrability. Also, which is not to be
underestimated, trying to use both descriptions immediately shows that some
objects are ill-defined and some can be defined in many non-equivalent ways.

To be more concrete, the variations or small perturbations to a given dynamical
system are an example of an object that reveals more when looked at from the
geometric point of view. It is fundamentally different from the original system
itself, as it really a vector on the tangent bundle of the main trajectory. It
also turns out that it should commute with the vector field defining the
system, as only then the transition to a nearby trajectory makes sense. It is
also the only additional field we need to reconstruct the congruence of
solutions if we can obtained the (first) variation as a function of the point on
the manifold. As it is usually not the case, and we only solve an equation that
gives the values of the variation on a particular trajectory, higher variations
are needed.

Here also the basic notions of differential geometry are helpful to prove the
existence of first integrals of the higher variational equations when the main
system has a first integral. Unfortunately it appears, that higher
variations, when defined to agree with the algebraic definitions, are not
coordinate independent. A fact hard to notice when analysing the equation only
in the coordinates in which it is introduced or obtained from physical
considerations.

Another example of clear formulation is the Lyapunov matrix and the Lyapunov
exponents. By definition they are constructed in a covariant way, but there is
a price to pay for that. Namely the additional metric structure is required.
Lyapunov exponents are usually computed using the time parameter that is
naturally present in system of physical origin, but in general relativistic
problems or those whose formulation admits the freedom of time
reparametrisation it is not clear which variable is the real time. And it is
obvious that a simple exponential change of that variable could make positive
exponents zero \cite{Marek_Lyap}.

The calculations presented here do not require any particular choice of metric,
so that they can be applied to any case and guarantee consistency. On the other
hand, without any particular choice it is impossible to obtain any results.
That is why the examples included are treated as is usually the case -- with
the tacit assumption that the coordinates in which the system is defined are
orthonormal. Until a distinguished metric structure can be canonically defined
for dynamical systems (or at least the physical systems), this freedom of
choice will remain unresolved.

As mentioned in the introduction there are attempts to geometrise the system
by finding some metric which would make the equations be the geodesic equations
on a suitable manifold, but so far this has been done for a small class of
systems with natural kinetic energy. It also immediately collides with the
problem of the base space -- in the case of Jacobi metric for example, only the
configuration space is taken into account, instead of the whole phase space.

This is best visible for Hamiltonian mechanics where we end up with analysing
second order equations in the coordinates and the momenta (although also
formally included in the solutions) do not play any role in the behaviour of
neighbouring trajectories. Because the Jacobi geometrisation hinges heavily on
the natural form of the kinetic energy, it is even impossible to obtain an
analogous picture with a space of half the dimension involving only the momenta
and suppressing the coordinates. The present work also shows that the symplectic
structure of such systems requires some serious additional metric assumptions
to speak about volume conservation. Even with the freedom that
differential geometry gives, Hamiltonian systems become highly structured in
this context.

Finally, among the still open problems, there is the question of studying more
than just the Levi-Civita connection for which the results reduce to the
algebraic ones. Introducing non-Riemannian (non-metric) connection or torsion,
complicates the equations considerably, but has, seemingly, nothing to do with
the question of integrability. This could hopefully give the possibility of
investigating the system on many different manifolds and in fact obtaining
different restrictions on integrability of the same basic equations.

Also, the algebraic tools are deeply rooted in the complex analysis of
meromorphic functions, Riemann surfaces and analytic continuation. Thus, being
integrable in the real sense is only understood indirectly. Here also lie new
possibilities of extending the work to complex or K\"ahler manifolds, or
developing the algebraic theory to treat the real-analytic case with more
detail.

\end{document}